\documentclass[preprint,12pt]{elsarticle}

\usepackage{lineno,hyperref}
\modulolinenumbers[5]

\journal{Information Sciences}

\usepackage{longtable}
\usepackage{subfigure}
\usepackage{url}
\usepackage{booktabs}
\usepackage{tabularx}
\usepackage{amsthm}
\usepackage{multirow}
\usepackage{colortbl}

\usepackage[final]{changes}
  \definechangesauthor[name={Tao Wen}, color=orange]{add}

  \usepackage{soul}
  \soulregister\cite7
  \soulregister\ref7

\newtheorem{definition}{Definition}[section]

\begin{document}

\begin{frontmatter}

\title{Identification of influencers in complex networks by local information dimensionality}

\author[address1]{Tao Wen}
\author[address1]{Yong Deng \corref{label1}}

\address[address1]{Institute of Fundamental and Frontier Science, University of Electronic Science and Technology of China, Chengdu, 610054, China}
\cortext[label1]{Corresponding author at: Institute of Fundamental and Frontier Science, University of Electronic Science and Technology of China, Chengdu, 610054, China. E-mail: dengentropy@uestc.edu.cn, prof.deng@hotmail.com.(Yong Deng)}
\begin{abstract}

\replaced{The identification of}{Identification} influential spreaders in \deleted{the} complex networks \replaced{is}{has been} a \replaced{popular}{hot} topic in \replaced{studies}{the study} of network \replaced{characteristics}{characteristic}. \replaced{Many}{Lots of} centrality measures have been proposed to address this problem, but most \deleted{methods} have \replaced{limitations}{their own limitations and shortcomings}. In this paper, a \deleted{novel} method \replaced{for}{is proposed to} \replaced{identifying}{identify} influencers in complex networks via the local information dimensionality \added{is proposed}. \replaced{The}{This} proposed method considers the local \replaced{structural properties}{structure property} around the central node\replaced{; therefore,}{, so} the scale of locality only increases to half of the maximum value of the shortest distance from the central node. Thus\replaced{, the}{this} proposed method considers the \replaced{quasilocal}{quasi-local} information and reduces the computational complexity. The information (number of nodes) in boxes is \replaced{described via the}{considered by}  Shannon entropy, which is more reasonable. \replaced{A node is}{The node would be} more influential when \replaced{its}{the} local information dimensionality \deleted{of this node} is higher. In order to show the effectiveness of \replaced{the}{this} proposed method, five existing centrality measures are used as comparison methods to rank influential nodes in six real-world complex networks. In addition, a \replaced{susceptible--infected}{Susceptible-Infected} (SI) model and Kendall's tau coefficient are applied to show the correlation between different methods. \replaced{E}{The e}xperiment results show the superiority of \replaced{the}{this} proposed method.

\end{abstract}


\end{frontmatter}


\section{Introduction}

Complex networks \replaced{are}{has been} a \replaced{popular topic that has}{hot research and}  attracted researchers' attention in many fields \cite{MAJHI2019100Chimera}\deleted{,} because \replaced{they}{it} can be used as a detailed model for many real-world complex systems \replaced{such as}{, like} brain network\added{s} \cite{small2012Gallos}, message network\added{s} \cite{Xu2017Optimal,Wang2017Model}, human lives \cite{Helbing2015Saving}\added{,} and social systems \cite{SHEIKHAHMADI2017517marketing,SHEIKHAHMADI2017517online}. \replaced{Many}{In recent years, many} structural properties of complex networks are affected by some special nodes, \replaced{e.g., the}{like} scale-free \cite{Matthew2014Perc}, self-similarity \cite{Rosenberg2017Maximal}\added{,} and \deleted{the} fractal \cite{wb2019} \replaced{properties}{property} of complex network\added{s} \cite{Rosenberg2017Minimal}. In order to measure networks' properties effectively, many studies have been \replaced{conducted}{launched} to find these nodes with special properties, \replaced{e.g.,}{for example,} finding the most similar node \cite{wentao2019similar}, identifying influential nodes \cite{Zareie2018hierarchical,SHEIKHAHMADI2015833Improving}, \added{and} predicting potential links \cite{XIE201937}. \replaced{In particular}{Particularly}, \deleted{the} nodes with \added{the ability to be} high\added{ly} influential \deleted{ability} in complex network\added{s have} gradually attracted researchers' attention\deleted{,} because they have \replaced{a greater}{more critical}  influence on \replaced{the networks' properties}{networks' property} and structure than most other nodes, \replaced{as demonstrated for}{like} predicting \added{a} time series by \added{a} visibility graph \cite{Fan2019timeseries}, predicting \added{a} link by similar nodes \cite{BU201941}, detecting \replaced{communities}{community} in social networks \cite{XIE201975}, measuring \added{the} network complexity \cite{Zhang2019Groups}, \added{and} dividing \added{the} network structure \cite{YANG2019121259method,yang2019network}.

In general, each network has a specific node importance ranking, and different identification methods consider different \replaced{structural}{structure} properties \replaced{of}{in} the network, which would give different ranking lists. \replaced{Many}{A lot of} centrality measures have been proposed to identify influential nodes\added{,} and they can be divided into three categories\cite{L2016Vital}\replaced{:}{, including} neighborhood-based\deleted{centralities}, path-based\deleted{centralities}, and iterative\added{-}refinement centralities. These centralities have many classical measures\deleted{,} such as \added{the} \replaced{degree centrality}{Degree Centrality} (DC) \cite{Newman2003Newman}, \replaced{betweenness centrality}{Betweenness Centrality} (BC) \cite{Newman2003Newman}, \replaced{closeness centrality}{Closeness Centrality} (CC) \cite{Freeman1979Centrality}, and \replaced{eigenvector centrality}{Eigenvector Centrality} (EC) \cite{Freeman1979Centrality}\replaced{. In addition to}{, and some}  new \deleted{centrality} measures\deleted{,} \replaced{such as the}{like} H-index \replaced{centrality}{Centrality} \cite{Zareie2019Centrality}, \added{those in} optimal percolation theory \cite{Ferraro2018Finding}\replaced{ and}{,} evidence theory \cite{MO2019121538evidence}, \added{the technique for order preference by similarity to the ideal solution (TOPSIS)} \cite{Zareie2018TOPSIS}\added{,} and other measures \cite{Zareie2019neighborhood,wang2018amodified,ZAREIE2019217interest}. These centrality measures have been applied in various fields\deleted{,} such as game theory \cite{wang2016statistical}, human cooperation \cite{Matja2017Statistical}, evolutionary games \cite{PERC2010109Coevolutionary}, \replaced{relevant website ranking}{ranking relevant websites} \cite{Zhang2019impact}, and \replaced{node}{affecting nodes'} synchronization \cite{Feng2018Synchronization,XING2019113}. However, these classical centrality measures have \replaced{limitations}{their own shortcomings and limitations}. For example, \added{the} DC only concentrates on \deleted{the} local information \replaced{and does not consider}{but lacks the consideration of} global information. \added{The} BC and CC \deleted{would} focus on \deleted{the} global information\added{,} but \replaced{their}{the} high computational complexity limits their application \replaced{to}{on} large-scale complex networks. \added{The} EC cannot be used in asymmetric networks, which \replaced{reduces}{would reduce} its application. Recently, some new centrality measures have been proposed. For instance, Zareie \emph{et al.} ranked \deleted{the} influential nodes \replaced{using the}{based} entropy \cite{Zareie2017Influential}. Deng \emph{et al.} proposed \added{a} local dimension to identify vital nodes \cite{Pu2014Identifying}. Makse \emph{et al.} traced \added{the} real information flow in social networks to find influential spreaders \cite{Teng2016Collective}.

\replaced{The entropy}{Entropy} is a useful tool \replaced{for measuring}{to measure} the information of \added{a} complex network\replaced{; hence,}{so} it has been \replaced{widely}{wildly} used \replaced{in many applications}{in many aspects in the network}, \added{e.g., evaluations of the} vulnerability \deleted{evaluating} \cite{wentao2018evaluating}, \added{presentation of the} dimension \deleted{presentation} \cite{Rosenberg2017Non,Pedrycz2003Fuzzy}, dilemma experiments \cite{Collective2012Gallos}, data fusion \cite{Song2019divergence,Gao2019generalizationn}, entanglement measures \cite{yang2018network}, and evidence theory \cite{Jiang2019Znetwork,liu2019new}. In addition, the structure of \replaced{a complex network}{complex networks}, such as \added{the} nodes and links, can be seen as probability sets. Therefore, the structural properties can be effectively explored by \added{the} entropy, which provides a new approach to address \replaced{problems}{the problem} in the network, including \added{the identification of} important nodes \deleted{identification}.

In this paper, a new centrality measure is proposed to identify \deleted{the} influential nodes \replaced{on the basis of the}{based on} local information dimensionality. \replaced{The}{This} proposed method considers the information in boxes through \added{the} Shannon entropy, which is more reasonable than classical measures. \replaced{In contrast to}{Different from} previous methods, the scale of locality \replaced{of the}{in this} proposed method \replaced{increases}{would grow} from \replaced{one}{1} to half of the maximum value of the shortest distance, which can consider the \replaced{quasilocal}{quasi-local} information and reduce the computational complexity. Nodes with \added{a} higher local information dimensionality are more influential in the complex network, which is \added{the} same as classical measures. \replaced{To}{In order to} show the effectiveness and reasonability of \replaced{the}{this} proposed method, six real-world complex networks are \replaced{considered}{used in this paper}, and five existing centrality measures are applied as comparison methods. \replaced{Further}{Fuethermore}, \replaced{a susceptible--infected}{Susceptible-Infected} (SI) model and Kendall's tau coefficient \cite{L2016Vital} are used to show the superiority of \replaced{the}{this} proposed method and the relationship between different methods.

The \deleted{organization of the} rest of this paper is \added{organized} as follows. Section 2 introduces some \replaced{existing}{exiting} centrality measures and concepts about complex networks. \replaced{The}{This} proposed local information dimensionality is \replaced{discussed}{proposed} in Section 3. \replaced{S}{Meanwhile, s}ome numerical experiments are \replaced{presented}{simulated} \added{in Section 4} to illustrate the effectiveness and reasonability of \replaced{the}{this} proposed method \deleted{in Section 4}. The conclusion\added{s} \replaced{are discussed}{is conducted} in Section 5.

\section{Preliminaries}

\subsection{\replaced{S}{The s}hortest distance between any two nodes in complex networks}

In a given complex network \emph{$G(N,V)$}, \emph{$N$} is the set of nodes, and \emph{$V$} is the set of edges. The adjacency matrix of the network can be obtained \replaced{from}{by} the topological \replaced{properties}{property} of network (the relationship\added{s} between \added{the} nodes and \added{the} edges)\replaced{. Then,}{, then} the shortest distance matrix can be obtained when the shortest distances between any two nodes are calculated by the Dijkstra algorithm. The adjacency matrix and shortest distance matrix are \added{the} known information of complex networks\deleted{,} and are solved in advance to facilitate later application. The shortest distance \emph{${\omega _{ij}}$} between node \emph{$i$} and node \emph{$j$} is defined as follows\replaced{:}{,}
\begin{equation}\label{equ_dis_min}
{\omega _{ij}} = \min ({e_{i{k_1}}} + {e_{{k_1}{k_2}}} +  \cdots  + {e_{{k_m}j}})
\end{equation}
where \emph{${k_1},{k_2}, \cdots ,{k_m}$} \replaced{are the}{is} node IDs and \emph{${e_{{k_1}{k_2}}}$} is the edge between two nodes. \emph{${e_{{k_1}{k_2}}} = 1$} \replaced{indicates that}{represents} there is an edge between two nodes, and \emph{${e_{{k_1}{k_2}}} = 0$} is the opposite. \replaced{Thus,}{So} the shortest \added{path} length \deleted{of path} between node \emph{$i$} and node \emph{$j$} is \replaced{denoted by}{represented as} \emph{${\omega _{ij}}$}, and the maximum value of the shortest distance from node \emph{$i$} is \deleted{shown below,}
\begin{equation}\label{equ_max_value_dis}
{\kappa _i} = \mathop {\max }\limits_{j \in N,j \ne i} ({\omega _{ij}})
\end{equation}
The maximum value of the shortest distance \emph{${\kappa _i}$} is the scale of locality around node \emph{$i$}, and it is different for different nodes.

\subsection{Centrality measures}

Some existing measures are introduced in this section \replaced{such as the BC, CC, DC, EC, and local dimension (LD).}{, like Betweenness Centrality (BC), Closeness Centrality (CC), Degree Centrality (DC), Eigenvector Centrality (EC), Local dimension (LD).}

\begin{definition}
Betweenness Centrality (BC) \cite{Newman2003Newman}. The \replaced{BC}{Betweenness Centrality} of node \emph{$i$} is denoted \replaced{by}{as} \emph{${C_B}(i)$} and defined as follows\replaced{:}{,}
\begin{equation}\label{equ_BC}
{C_B}(i) = \sum\limits_{s,t \ne i} {\frac{{{g_{st}}(i)}}{{{g_{st}}}}}
\end{equation}
where \emph{${{g_{st}}}$} is the number of shortest paths between node \emph{$s$} and node \emph{$t$}\deleted{,} and \emph{${{g_{st}}(i)}$} is the number of shortest paths between node \emph{$s$} and node \emph{$t$} \replaced{that pass}{which go} through node \emph{$i$}.
\end{definition}

\begin{definition}
Closeness Centrality (CC) \cite{Freeman1979Centrality}. The \replaced{CC}{Closeness Centrality} of node \emph{$i$} is denoted \replaced{by}{as} \emph{${C_C}(i)$} and defined as follows\replaced{:}{,}
\begin{equation}\label{equ_CC}
{C_C}(i) = {\left( {\sum\limits_{j = 1}^{\left| N \right|} {{\omega _{ij}}} } \right)^{ - 1}}
\end{equation}
where \emph{${\omega _{ij}}$} is the shortest distance from node \emph{$i$} \replaced{to}{and} node \emph{$j$} \replaced{that}{which} can be obtained by Eq. (\ref{equ_dis_min})\deleted{,} and \emph{$\left| N \right|$} is the number of nodes.
\end{definition}

\begin{definition}
Degree Centrality (DC) \cite{Newman2003Newman}. The \replaced{DC}{Degree Centrality} of node \emph{$i$} is denoted \replaced{by}{as} \emph{${C_D}(i)$} and defined as follows\replaced{:}{,}
\begin{equation}\label{equ_DC}
{C_D}(i) = \sum\limits_{j = 1}^{\left| N \right|} {{e_{ij}}}
\end{equation}
where \emph{${{e_{ij}}}$} \replaced{is}{shows} the edge between node \emph{$i$} and \emph{$j$}\deleted{, and \emph{$\left| N \right|$} is the number of nodes.}. In fact, the \replaced{DC}{Degree Centrality} means the number of edges connected with the selected node.
\end{definition}

\begin{definition}
Eigenvector Centrality (EC) \cite{Freeman1979Centrality}. \emph{$A$} is a similarity matrix whose size is \emph{$\left| N \right| \times \left| N \right|$}. The \replaced{EC}{Eigenvector Centrality} \emph{${x_i}$} of node \emph{$i$} is the \emph{$i$}th entry in the normalized eigenvector \replaced{that}{which} belongs to \emph{$A$}, and it is defined as follows\replaced{:}{,}
\begin{equation}\label{equ_EC}
Ax = \lambda x,{x_i} = u\sum\limits_{j = 1}^{\left| N \right|} {{a_{ij}}{x_j}}
\end{equation}
where \emph{$\lambda$} is the largest eigenvalue of \emph{$A$}, \deleted{and} \emph{$u = 1/\lambda $}, \deleted{\emph{$\left| N \right|$} is the number of nodes,} \added{and} \emph{${x_i}$} is the sum of \added{the} similarity scores of the nodes \replaced{that}{which} are connected with node \emph{$i$}.
\end{definition}

The \replaced{LD}{local dimension} \cite{Pu2014Identifying} of node \emph{$i$} is introduced in Section 2.3.

\subsection{Local dimension}

To explore the local structural properties of complex networks, Silva \emph{et al.} proposed the \replaced{LD}{local dimension} of complex networks. \replaced{A}{The} power-law distribution has been \replaced{proven}{proved} to exist in theoretical networks with special properties \replaced{such as}{like} small-world \replaced{properties and}{, but also in} many real-world networks. Because the topological scale from each central node is different, the \replaced{LD}{local dimension} \replaced{changes}{would change} with the selection of the central node. \deleted{Then,} Pu \emph{et al.} \cite{Pu2014Identifying} modified \replaced{the LD}{local dimension} to identify the vital nodes in complex networks. For a radius \emph{$r$}\replaced{, it has been found that}{and} the number of nodes \emph{${N_i}(r)$} whose shortest distance from \added{the} central node \added{is} less than \emph{$r$} follows \replaced{the}{a} power law\deleted{, and it is shown as follows,}
\begin{equation}\label{equ_power_law}
{N_i}(r) \sim {r^{{D_i}}}
\end{equation}
It can be easily found that the \replaced{LD}{local dimension} \emph{${D_i}$} of node \emph{$i$} can be obtained \replaced{from}{by} the slope of \replaced{a log--log plot}{double logarithmic curves}, and it is \replaced{expressed}{shown} as follow\replaced{s:}{,}
\begin{equation}\label{equ_local_dimen}
{D_i} = \frac{d}{{d\ln r}}\ln {N_i}(r)
\end{equation}
where \emph{$d$} is the symbol of derivative. The radius \emph{$r$} \replaced{increases}{would grow} from \replaced{one}{1} to the maximum value of the shortest distance \emph{${\kappa_i}$} from node \emph{$i$}, and the derivative of Eq. (\ref{equ_local_dimen}) \replaced{is expressed as follows}{can be shown below} because of the discrete properties \cite{Ben2004Complex} \replaced{of}{in} complex networks\replaced{:}{,}
\begin{equation}\label{equ_local_dimen_deri_1}
{D_i} = \frac{r}{{{N_i}(r)}}\frac{d}{{dr}}{N_i}(r)
\end{equation}
\begin{equation}\label{equ_local_dimen_deri_2}
{D_i} = r\frac{{{n_i}(r)}}{{{N_i}(r)}}
\end{equation}
where \emph{${{n_i}(r)}$} is the number of nodes whose shortest distance from \added{the} central node equal\added{s} \emph{$r$}. When a central node is chosen, the \deleted{locality} scale \added{of locality} of the central node \replaced{can}{would} be determined, and the \replaced{LD}{local dimension} of \added{the} central node can be obtained \replaced{from}{by} the slope of \replaced{a log--log plot}{double logarithmic curves} (\emph{$\ln {N_i}(r)$} vs. \emph{$\ln r$}). Lastly, the importance of \added{a} node can be determined by the order of the \replaced{LD}{local dimension}. \replaced{In contrast to}{Different from} the previous method\added{s}, \replaced{a}{the} node with a lower \replaced{LD is}{local dimension would be} more influential in the network.

\section{\replaced{P}{The p}roposed method}

\begin{figure}
\centering
\includegraphics[width=15cm]{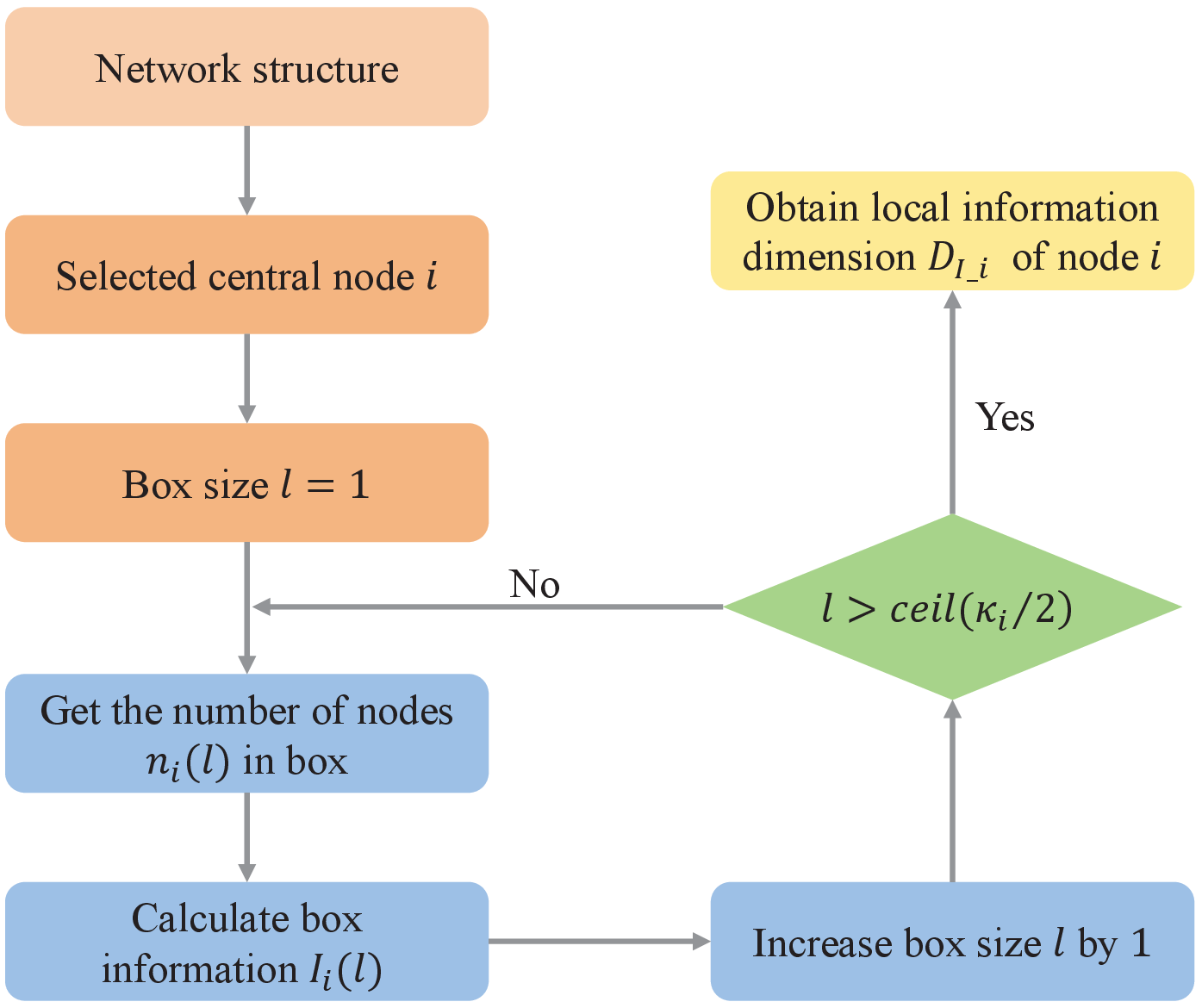}\\
\caption{\textbf{\replaced{F}{The f}lowchart of \replaced{the}{this} proposed method.} The main step \added{is} to calculate \replaced{the}{one node's} local information dimensionality \emph{${D_{I\_i}}$} \added{of one node} from the structure of \added{a} complex network.}
\label{fig_flow chart}
\end{figure}

\replaced{Many}{Lots of} centrality measures have been proposed to identify the influential nodes in \deleted{the} complex network\added{s}. Different methods consider different \replaced{structural}{structure} information in the network and \deleted{they would} have their own advantages and limitations. Because most \deleted{of the} previous methods concentrate on the global \deleted{structure} or local structure, the \replaced{quasilocal}{quasi-local} structure around the selected node cannot be \added{effectively} recognized\deleted{effectively}. In this paper, a new method is proposed to identify vital nodes \replaced{on the basis of}{based on} the local information dimensionality (LID) of each node in \added{a} complex network. \replaced{The}{This} proposed method \replaced{considers}{would consider} the \replaced{quasilocal}{quasi-local} information around each node and reduce\added{s} the computational complexity. The \replaced{practicality}{practicability} and effectiveness of \replaced{the}{this} proposed method \replaced{are demonstrated}{can be shown} \replaced{with experiments comparing}{from comparison experiments in} some real-world complex networks in Section 4. \replaced{A}{The} flowchart \replaced{for}{to} \replaced{obtaining}{obtain} the \replaced{LID}{local information dimension} of one selected node is shown in Fig. \ref{fig_flow chart}.

In this section, the \replaced{LID}{local information dimension} of \added{a} complex network is proposed. The number of nodes in each box \replaced{is}{would be} considered \replaced{using the}{by} Shannon entropy in \replaced{the}{this} proposed method, which is more reasonable. \replaced{Similar to the LD}{The same as local dimension} \emph{${D_i}$}, the \replaced{LID}{local information dimension} \emph{${D_{I\_i}}$} also considers the \replaced{structural}{structure} properties around node \emph{$i$} in complex networks, and it is defined as follows\replaced{:}{,}
\begin{equation}\label{equ_LID}
{D_{I\_i}} =  - \frac{d}{{d\ln l}}{I_i}(l)
\end{equation}
where \emph{$d$} is the symbol of derivative, \emph{$l$} is the size of the box\deleted{,} and \emph{${I_i}(l)$} \replaced{is}{represents} the information in the box whose central node is node \emph{$i$} with size \emph{$l$}. \replaced{In contrast to the classical LD}{Different from the classical local dimension}, the information \emph{${I_i}(l)$} in the selected box is considered \replaced{using the}{by} Shannon entropy to describe the number of nodes in the box. In addition, the \added{rule that governs the growth of the size of the} box \deleted{size growth rule} is different from the classical definition. The size of \added{the} box \emph{$l$} \replaced{grows}{would grow} from \replaced{one}{1} to \deleted{the} half of the maximum value of the shortest distance from \added{the} central node \emph{${\kappa _i}$}, \added{i.e.,} \emph{$ceil({\kappa _i}/2)$}. The change \replaced{in the size of the}{of} box \deleted{size} means \added{that the} LID \replaced{focuses}{would focus} on the \replaced{quasilocal}{quasi-local} structure around the central node and \replaced{reduces the}{reduce} computational complexity. The information \emph{${I_i}(l)$} in each box is determined by the number of nodes in the box, and the number of \deleted{the} selected nodes is considered \replaced{using the}{by} Shannon entropy. Thus, the information \replaced{in the}{of} box can \replaced{indicate}{show} the node's properties more \replaced{reasonably}{reasonable}, and it is \deleted{detailed} defined as follows\replaced{:}{,}
\begin{equation}\label{equ_I_i}
{I_i}(l) =  - {p_i}(l)\ln {p_i}(l)
\end{equation}
where \emph{${p_i}(l)$} is the probability \replaced{that}{of} information \replaced{is contained}{containing} in \replaced{a}{the} box whose central node is \emph{$i$} for a given box size \emph{$l$}, which is the ratio of the number of nodes in the box\added{, \emph{${{n_i}(l)}$},} to the \replaced{total}{whole} number \added{of nodes} in the complex network\added{, \emph{$N$},} and can be obtained as follows\replaced{:}{,}
\begin{equation}\label{equ_p_i}
{p_i}(l) = \frac{{{n_i}(l)}}{N}
\end{equation}
\deleted{where \emph{${{n_i}(l)}$} is the number of nodes within the box whose size equals to \emph{$l$}, and \emph{$N$} is the whole number of nodes in the complex network.} Thus, \replaced{the LID}{this proposed local information dimension shown} in Eq. (\ref{equ_LID}) can be rewritten as follows\replaced{:}{,}
\begin{equation}\label{equ_LID_new}
{D_{I\_i}} =  - \frac{d}{{d\ln l}}\left( { - \frac{{{n_i}(l)}}{N}\ln \frac{{{n_i}(l)}}{N}} \right)
\end{equation}
\replaced{From}{It can be obtained from} Eq. (\ref{equ_LID_new})\replaced{,}{that} the information \replaced{in the}{of} box\added{,} \emph{${I_i}(l)$}\added{,} around the central node \emph{$i$} is obtained from the number of nodes in the box by \added{the} Shannon entropy. The \replaced{LID}{locla information dimension} \emph{${D_{I\_i}}$} of \added{the} selected node \emph{$i$} is obtained \replaced{from}{by} the slope of the \deleted{fitting} line \added{fitting the relationship} between \emph{${I_i}(l)$} and \emph{${\ln l}$}. Because of \added{the} network\replaced{'s}{s'} discrete nature \cite{Ben2004Complex}, the \added{expression with the} derivative \deleted{expression} in Eq. (\ref{equ_LID_new}) can be rewritten \replaced{as}{below,}
\begin{equation}\label{equ_LID_derivative}
\begin{array}{l}
{D_{I\_i}} =  - \frac{d}{{d\ln l}}\left( { - {p_i}(l)\ln {p_i}(l)} \right)\\
{\rm{       }} = \frac{l}{{1 + \ln {p_i}(l)}}\frac{d}{{dl}}{p_i}(l)\\
{\rm{       }} \approx \frac{l}{{1 + \ln \frac{{{n_i}(l)}}{N}}}\frac{{{N_i}(l)}}{N}
\end{array}
\end{equation}
where \emph{${{N_i}(l)}$} is the number of nodes whose shortest distance from \added{the} central node \emph{$i$} equals \replaced{the}{to} box size \emph{$l$} (\emph{${\omega _{ij}} = l$})\deleted{,} and \emph{${{n_i}(l)}$} is the number of nodes whose shortest distance from \added{the} central node \emph{$i$} is less than the box size \emph{$l$} (\emph{${\omega _{ij}} \le l$}).

In \replaced{the}{this} proposed method, the scale of locality \emph{${r_{\max }}$} \replaced{changes}{would change} with the central nodes, which is defined as the half of the maximum value of the shortest distance from \added{the} central node \emph{${\kappa_i}$}, i.e.\added{,} \emph{${r_{\max }} = ceil({\kappa_i}/2)$}. The box size \emph{$l$} \replaced{increases}{would increase} from \replaced{one}{1} to the scale of locality \emph{${r_{\max }}$}. The information in each box (\added{the} number of nodes in the box) \replaced{is}{would be} considered \replaced{using the}{by} Shannon entropy. The \replaced{LID}{local information dimension} of each node can be obtained \replaced{from}{by} the slope of \added{the} box information \emph{${I_i}(l)$} and the logarithm of the box size \emph{${\ln l}$}. \replaced{Owing}{Due} to the \replaced{properties}{property} of the \replaced{LID}{local information dimension}, \replaced{the}{this} proposed method \replaced{considers}{would consider} the information in the box more \replaced{reasonably}{reasonable} and \replaced{reduces}{reduce} the computational complexity.

%
%
%
%

\section{Experimental study}

\replaced{To}{In order to} show the effectiveness of \replaced{the}{this} proposed method, six real-world complex networks and five \replaced{comparison}{comparing} measures are used \deleted{in this section}. These six complex networks are \added{the} USAir network, Jazz network, Karate network, Political blogs network, Facebook network, and \replaced{(High Energy Physics - Theory) collaboration network from arXiv}{Collaboration network of Arxiv high energy physics theory respectively}, which can be downloaded from \deleted{(}$http://vlado.fmf.uni-lj.si/pub/networks/data/$\deleted{)} and \deleted{(}$http://snap.stanford.edu/data/$\deleted{)}. In addition, the collaboration network is chosen as the largest connected subgraph from \added{the} original network data. These five comparison measures (\added{the} BC, CC, DC, EC, \added{and} LD) \replaced{were}{have been} introduced in Section 2. The structural properties of these six networks are \replaced{listed}{shown} in Table \ref{table_Property}. $\left| N \right|$ and $\left| V \right|$ \replaced{are}{represent} the numbers of nodes and edges\deleted{,} respectively\replaced{;}{,} $<k>$ and ${{k_{max}}}$ \replaced{are}{mean} the average and maximum value of \added{the} degree\added{of centrality, respectively; and} $<\omega>$ and ${{\omega_{max}}}$ \replaced{are}{mean} the average and maximum value of the shortest distance in the network\added{, respectively}. Five \deleted{kinds of} experiments are implemented \deleted{in this section}, including listing the top-10 node IDs to compare the difference\added{s} \replaced{in the}{about} top-10 node results obtained by different measures\replaced{,}{;} \replaced{the propagation}{propagating} based on \added{the} SI model to show the superior \replaced{infection}{infectious} ability of \added{the} nodes obtained by \added{the} LID\replaced{, the}{;} \added{the} relationship graph and Kendall's tau coefficient to show the similarity of the node rankings obtained by different measures and \added{the} SI model\replaced{, and the}{;} running time\added{s} of different measures to show \replaced{the}{this} proposed method's low computational complexity.

\begin{table}[!htbp]
\centering
\caption{\textbf{\replaced{Topological properties of r}{R}eal-world network\replaced{s}{ topological properties}.}}
\begin{tabular}{ccccccc}
\hline
Network & $\left| N \right|$  & $\left| V \right|$  & $<k>$  & ${{k_{max}}}$  & $<\omega>$  & ${{\omega_{max}}}$ \\
\hline
USAir              & 332 & 2126 & 12.8072 & 139 & 2.7381 & 6\\
Jazz               & 198 & 5484 & 27.6970 & 100 & 2.2350 & 6\\
Karate             & 34  & 78   & 4.5882  & 17  & 2.4082 & 5 \\
Political blogs    &1222 &19021 & 27.3552 & 351 & 2.7375 & 8 \\
Facebook           &4039 &88234 & 43.6910 &1045 & 3.6925 & 8\\
Collaboration      &8368 &24827 & 5.7459  & 65  & 5.9454 & 18\\
\hline
\end{tabular}
\label{table_Property}
\end{table}

\subsection{Top-10 nodes}

First\added{ly}, the top-10 nodes in six real-world complex networks are identified by \added{the} LID and five other \deleted{exiting} centrality measures, and the results are \replaced{listed}{shown} in Table \ref{table_top_10}. \replaced{The}{These} nodes \replaced{in}{with} color \replaced{for the}{in} five existing measures \deleted{show that they} are the same top-10 nodes \replaced{identified by the}{in} LID. Because of \added{the} different consideration of information in the network, different centrality measures \added{could} give different \added{lists of} top-10 node\added{s} \deleted{lists}. Thus, the numbers of same nodes \replaced{for}{between} different measures can show the similarity of information considered by different methods, and \replaced{the identification of more nodes by the}{more nodes in} LID\added{,} which also appear in the results of other measures\added{,} can \replaced{increase its}{bring more} credibility \deleted{to LID}. These unique nodes \replaced{identified by the}{in} LID may \replaced{indicate}{have} \deleted{an} important change\added{s} \replaced{in}{to} the propagation process. \replaced{From}{Observing from} Table \ref{table_top_10}, \deleted{in USAir network,} the \added{same node with the} most influence\deleted{r} \added{in the USAir network is} \deleted{node} obtained by \added{the} six different methods \replaced{---}{are the same, and it is}node 118. The top-10 nodes obtained by \added{the} LID \replaced{are}{is} the same as \added{those obtained by the} CC, and \deleted{there are also} eight and seven \added{of the} same nodes \added{are obtained} between \added{the} LID and \added{the} DC \replaced{and}{,} EC\added{,} respectively. The number\added{s} of \added{the} same top-10 nodes obtained by \added{the} BC, LD\added{,} and LID are less than \added{those of} other methods\added{,} which are \replaced{six}{only six nodes}. The result\added{s} \replaced{for the}{in} USAir network show that most of \added{the} top-10 nodes \replaced{obtained by the}{in} other measures \replaced{are obtained by the}{appear in} LID, which \replaced{indicates its}{shows the} similarity and credibility\deleted{ of this proposed method}.

\replaced{For the}{In} Jazz network, the number of same top-10 nodes between \added{the} LID and \added{the} other measures is relatively small. There are only five same top-10 nodes \replaced{among the}{between} CC, DC, EC\added{,} and LID\replaced{;}{,} \replaced{the BC and LD}{other methods (BC, LD)} only have \replaced{three}{3} and \replaced{four of the}{4} same nodes\added{, respectively,} \replaced{as the}{with} LID\deleted{respectively}. The difference\added{s} \replaced{in the}{of} top-10 nodes between \added{the} LID and \added{the} other measures are \replaced{large}{big}, but \replaced{the}{these} unique nodes in \added{the} LID \deleted{would} have a \replaced{significant}{big} influence on the propagation process\added{,} which can show \replaced{the}{these nodes'} importance \added{of these nodes} in complex networks. \replaced{A}{The} detailed comparison \replaced{is}{would be} carried out \replaced{with}{in} the experiments below.

\replaced{The}{Observing from the} \added{list of} top-10 nodes \replaced{of the}{list of} Karate network in Table \ref{table_top_10}, \deleted{the top-10 nodes list} is exactly the same using \added{the} LID and CC (\replaced{this}{This} result is the same as the result \replaced{for the}{in} USAir network). \replaced{Many of the}{There are also many} same top-10 nodes \replaced{are observed for the}{in} DC, EC\added{,} and LID, \replaced{which}{and both of them} have \added{the} same \replaced{nine}{9} nodes, \replaced{thereby indicating}{which can show} the similarity between them. The number\added{s} of same top-10 nodes between \added{the} LID and \replaced{the BC and LD}{other methods (BC, LD)}  \replaced{are seven and five,}{is 7, 5} respectively.

\replaced{For the}{In} Political blogs network, the number\added{s} of same top-10 nodes \replaced{between the}{obtained by} LID and \added{the} CC, DC, \added{and} LD are 9, 9, \added{and} 10 respectively, which are \replaced{larger}{bigger} than the number\added{s} of same top-10 nodes obtained by \added{the} BC (\replaced{seven}{7 same nodes}) and EC (\replaced{five}{5 same nodes}).

\replaced{For the}{The, in} Facebook network, the \replaced{largest}{biggest} number of same top-10 nodes \added{as the LID} is \added{obtained for the} CC (\replaced{seven}{7 same nodes}), which demonstrates the similarity of information considered by these two measures in this network. The results \replaced{for the}{in} BC, DC, \added{and} LD are \replaced{five, four, and four,}{a little less, and is 5, 4, 4} respectively. \replaced{However}{But}, there are \replaced{no}{not the} same top-10 nodes between \added{the} EC and \added{the} LID, which \replaced{indicates that there is}{shows} a \replaced{large}{huge} difference between these two measures.

Lastly, \replaced{for}{in} the \replaced{largest}{biggest} network \replaced{---the collaboration}{named as Collaboration} network, the number of same top-10 nodes between \added{the} CC, LD\added{,} and LID \replaced{is}{are} five, which \replaced{is larger}{are relatively bigger} than \added{those for} other methods\deleted{' results}. \added{The number of same top-10 nodes between the LID and the} BC and EC \replaced{are significantly different:}{give big different results from LID, and they are} \replaced{one and zero,}{1, 0} respectively.

Thus, these centrality measures \replaced{consider the}{would have different consideration of} information in this network \replaced{differently}{,} and give different \added{lists of} top-10 nodes \deleted{lists}. \added{The} DC has four \added{of the} same top-10 nodes \replaced{as the}{with} LID, which is better than \added{the} BC and EC. \replaced{From}{Obtained from} Table \ref{table_top_10} and \replaced{the discussion above}{above discussion}, \added{the} CC and DC are two \deleted{similar} measures \replaced{similar to the}{with} LID\deleted{,} because they can \replaced{obtain}{get} closer \replaced{rankings}{ranks} than other methods. \added{The} BC and EC have different performance in these networks\replaced{;}{,} some networks have similar \replaced{rankings,}{ranks} and some are \replaced{different}{opposite}. The reason why \added{there are} no same top-10 nodes between \added{the} EC and \added{the} LID is the influence of \added{the} network scale. \added{The} EC does not have \deleted{a} good performance in large-scale network\added{s} because of its complete\added{ly} different \added{list of} top-10 node\added{s} \replaced{from}{list with} other methods. In conclusion, \replaced{the LID}{this proposed method (LID)} \replaced{exhibits}{has a close} performance \replaced{closer to}{the} existing measures \replaced{for}{on} identifying top-10 nodes\replaced{.}{,} \replaced{A}{and} more detail\added{ed} comparison \added{with} experiments \replaced{is}{are} \replaced{discussed}{illustrated} below. Because \replaced{the}{this} proposed method is modified from \added{the} LD\deleted{,} and it focuses on \replaced{a}{the} node's influence from different distances, the \deleted{main comparison measures are chosen as} LD (the most related method) and CC (the same consideration factor) \added{are chosen for comparison} in the experiments below.

\begin{table*}[!htbp]
\scriptsize
\centering
\caption{\textbf{\deleted{The} \replaced{T}{t}op-10 nodes ranked by different centrality methods in six real-world complex networks.} \replaced{A}{The} node \replaced{in}{with} color \replaced{indicates}{shows} that \replaced{it}{this node} also exists in the top-10 list \replaced{obtained by the}{of} LID. The \replaced{similarities in the}{similarity of} top-10 nodes between different measures and \added{the} LID are \replaced{provided}{shown}.}

\begin{tabular}{ccccccccccccccccccc}
\toprule
\multirow{2}{*}{Rank} & \multicolumn{6}{c}{USAir network} & \multicolumn{6}{c}{Jazz network} \\
\cmidrule(r){2-7} \cmidrule(r){8-13} \cmidrule(r){14-19}
& BC & CC & DC & EC & LD & LID & BC & CC & DC & EC & LD & LID  \\
\midrule
1 &\textcolor[rgb]{0,0,1}{118} & \textcolor[rgb]{1,0,0}{118} & \textcolor[rgb]{0,1,0}{118} & \textcolor[rgb]{0.3,0.3,0.8}{118} & \textcolor[rgb]{0.8,0.3,0.3}{118} & 118
&  \textcolor[rgb]{0,0,1}{136} & \textcolor[rgb]{1,0,0}{136} & \textcolor[rgb]{0,1,0}{136} & 60                                & 60                                & 136  \\
2 & 8                          & \textcolor[rgb]{1,0,0}{261} & \textcolor[rgb]{0,1,0}{261} & \textcolor[rgb]{0.3,0.3,0.8}{261} & \textcolor[rgb]{0.8,0.3,0.3}{261} & 67
&   60                         & 60                          & 60                          & \textcolor[rgb]{0.3,0.3,0.8}{132} & \textcolor[rgb]{0.8,0.3,0.3}{136} & 168\\
3 &\textcolor[rgb]{0,0,1}{261} & \textcolor[rgb]{1,0,0}{67}  & \textcolor[rgb]{0,1,0}{255} & \textcolor[rgb]{0.3,0.3,0.8}{255} & 152                               & 261
&  \textcolor[rgb]{0,0,1}{153} & \textcolor[rgb]{1,0,0}{168} & \textcolor[rgb]{0,1,0}{132} & \textcolor[rgb]{0.3,0.3,0.8}{136} & \textcolor[rgb]{0.8,0.3,0.3}{132} &70\\
4 &\textcolor[rgb]{0,0,1}{47}  & \textcolor[rgb]{1,0,0}{255} & \textcolor[rgb]{0,1,0}{182} & \textcolor[rgb]{0.3,0.3,0.8}{182} & 230                               & 201
&   5                          & \textcolor[rgb]{1,0,0}{70}  & \textcolor[rgb]{0,1,0}{168} & \textcolor[rgb]{0.3,0.3,0.8}{168} & \textcolor[rgb]{0.8,0.3,0.3}{83}  &122\\
5 &\textcolor[rgb]{0,0,1}{201} & \textcolor[rgb]{1,0,0}{201} & 152                         & 152                               & \textcolor[rgb]{0.8,0.3,0.3}{255} & 47
&   149                        & \textcolor[rgb]{1,0,0}{83}  & \textcolor[rgb]{0,1,0}{70}  & 108                               & \textcolor[rgb]{0.8,0.3,0.3}{168} &178\\
6 &\textcolor[rgb]{0,0,1}{67}  & \textcolor[rgb]{1,0,0}{182} & 230                         & 230                               & \textcolor[rgb]{0.8,0.3,0.3}{182} & 255
&   189                        & \textcolor[rgb]{1,0,0}{132} & 108                         & 99                                & 99                                &83\\
7 & 313                        & \textcolor[rgb]{1,0,0}{47}  & \textcolor[rgb]{0,1,0}{166} & \textcolor[rgb]{0.3,0.3,0.8}{112} & \textcolor[rgb]{0.8,0.3,0.3}{112} & 166
&   167                        & 194                         & 99                          & 131                               & 108                               &18\\
8 & 13                         & \textcolor[rgb]{1,0,0}{248} & \textcolor[rgb]{0,1,0}{67}  & \textcolor[rgb]{0.3,0.3,0.8}{67}  & 147                               & 248
&   96                         & \textcolor[rgb]{1,0,0}{122} & 158                         & \textcolor[rgb]{0.3,0.3,0.8}{70}  & 158                               &153\\
9 &\textcolor[rgb]{0,0,1}{182} & \textcolor[rgb]{1,0,0}{166} & \textcolor[rgb]{0,1,0}{112} & \textcolor[rgb]{0.3,0.3,0.8}{166} & \textcolor[rgb]{0.8,0.3,0.3}{166} & 182
&   115                        & 174                         & \textcolor[rgb]{0,1,0}{83}  & \textcolor[rgb]{0.3,0.3,0.8}{83}  & 194                               &118\\
10 & 152                       & \textcolor[rgb]{1,0,0}{112} & \textcolor[rgb]{0,1,0}{201} & 147                               & 293                               & 112
&  \textcolor[rgb]{0,0,1}{83}  & 158                         & 7                           & 194                               & 7                                 &132\\

\bottomrule

\multirow{2}{*}{Rank} & \multicolumn{6}{c}{Karate network} & \multicolumn{6}{c}{Political blogs network} \\
\cmidrule(r){2-7} \cmidrule(r){8-13} \cmidrule(r){14-19}
& BC & CC & DC & EC & LD & LID & BC & CC & DC & EC & LD & LID  \\
\midrule
1  & \textcolor[rgb]{0,0,1}{1}  & \textcolor[rgb]{1,0,0}{1}  & \textcolor[rgb]{0,1,0}{34} & \textcolor[rgb]{0.3,0.3,0.8}{34} & \textcolor[rgb]{0.8,0.3,0.3}{34} & 32
   & \textcolor[rgb]{0,0,1}{12} & \textcolor[rgb]{1,0,0}{28} & \textcolor[rgb]{0,1,0}{12} & \textcolor[rgb]{0.3,0.3,0.8}{12} & \textcolor[rgb]{0.8,0.3,0.3}{12} & 12  \\
2  & \textcolor[rgb]{0,0,1}{3}  & \textcolor[rgb]{1,0,0}{3}  & \textcolor[rgb]{0,1,0}{1}  & \textcolor[rgb]{0.3,0.3,0.8}{1}  & \textcolor[rgb]{0.8,0.3,0.3}{1}  & 3
   & \textcolor[rgb]{0,0,1}{304}& \textcolor[rgb]{1,0,0}{12} & \textcolor[rgb]{0,1,0}{28} & \textcolor[rgb]{0.3,0.3,0.8}{14} & \textcolor[rgb]{0.8,0.3,0.3}{28} & 28  \\
3  & \textcolor[rgb]{0,0,1}{34} & \textcolor[rgb]{1,0,0}{34} & \textcolor[rgb]{0,1,0}{33} & \textcolor[rgb]{0.3,0.3,0.8}{3}  & \textcolor[rgb]{0.8,0.3,0.3}{33} & 14
   & \textcolor[rgb]{0,0,1}{94} & \textcolor[rgb]{1,0,0}{16} & \textcolor[rgb]{0,1,0}{304}& \textcolor[rgb]{0.3,0.3,0.8}{16} & \textcolor[rgb]{0.8,0.3,0.3}{304}& 16  \\
4  & \textcolor[rgb]{0,0,1}{33} & \textcolor[rgb]{1,0,0}{32} & \textcolor[rgb]{0,1,0}{3}  & \textcolor[rgb]{0.3,0.3,0.8}{33} & 24                               & 9
   & \textcolor[rgb]{0,0,1}{28} & \textcolor[rgb]{1,0,0}{14} & \textcolor[rgb]{0,1,0}{14} & \textcolor[rgb]{0.3,0.3,0.8}{67} & \textcolor[rgb]{0.8,0.3,0.3}{14} & 14  \\
5  & \textcolor[rgb]{0,0,1}{32} & \textcolor[rgb]{1,0,0}{33} & \textcolor[rgb]{0,1,0}{2}  & \textcolor[rgb]{0.3,0.3,0.8}{2}  & \textcolor[rgb]{0.8,0.3,0.3}{3}  & 20
   & 145                        & \textcolor[rgb]{1,0,0}{36} & \textcolor[rgb]{0,1,0}{16} & 52                               & \textcolor[rgb]{0.8,0.3,0.3}{16} & 304 \\
6  & 6                          & \textcolor[rgb]{1,0,0}{14} & \textcolor[rgb]{0,1,0}{32} & \textcolor[rgb]{0.3,0.3,0.8}{9}  & \textcolor[rgb]{0.8,0.3,0.3}{2}  & 33
   & \textcolor[rgb]{0,0,1}{6}  & \textcolor[rgb]{1,0,0}{67} & \textcolor[rgb]{0,1,0}{94} & 18                               & \textcolor[rgb]{0.8,0.3,0.3}{94} & 94  \\
7  & \textcolor[rgb]{0,0,1}{2}  & \textcolor[rgb]{1,0,0}{9}  & \textcolor[rgb]{0,1,0}{4}  & \textcolor[rgb]{0.3,0.3,0.8}{14} & 30                               & 1
   & \textcolor[rgb]{0,0,1}{16} & \textcolor[rgb]{1,0,0}{94} & \textcolor[rgb]{0,1,0}{6}  & \textcolor[rgb]{0.3,0.3,0.8}{28} & \textcolor[rgb]{0.8,0.3,0.3}{6}  & 67  \\
8  & 28                         & \textcolor[rgb]{1,0,0}{20} &  24                        & \textcolor[rgb]{0.3,0.3,0.8}{4}  & 6                                & 2
   & 300                        & \textcolor[rgb]{1,0,0}{35} & \textcolor[rgb]{0,1,0}{67} & 47                               & \textcolor[rgb]{0.8,0.3,0.3}{67} & 36  \\
9  & 24                         & \textcolor[rgb]{1,0,0}{2}  & \textcolor[rgb]{0,1,0}{14} & \textcolor[rgb]{0.3,0.3,0.8}{32} & 7                                & 34
   & 163                        & 145                        & \textcolor[rgb]{0,1,0}{35} & 73                               & \textcolor[rgb]{0.8,0.3,0.3}{35} & 35  \\
10 & \textcolor[rgb]{0,0,1}{9}  & \textcolor[rgb]{1,0,0}{4}  & \textcolor[rgb]{0,1,0}{9}  & 31                               & 28                               & 4
   & \textcolor[rgb]{0,0,1}{35} & \textcolor[rgb]{1,0,0}{304}& 145                        & 9                                & \textcolor[rgb]{0.8,0.3,0.3}{36} & 6   \\

\bottomrule

\multirow{2}{*}{Rank} & \multicolumn{6}{c}{Facebook network} & \multicolumn{6}{c}{Collaboration network} \\
\cmidrule(r){2-7} \cmidrule(r){8-13} \cmidrule(r){14-19}
& BC & CC & DC & EC & LD & LID & BC & CC & DC & EC & LD & LID  \\
\midrule
1  & \textcolor[rgb]{0,0,1}{107}  & \textcolor[rgb]{1,0,0}{107}  & \textcolor[rgb]{0,1,0}{107}  & 801                        & \textcolor[rgb]{0.8,0.3,0.3}{107}  & 107
   & 3814                         & \textcolor[rgb]{1,0,0}{2448} & 187                          & 4809                       & \textcolor[rgb]{0.8,0.3,0.3}{1763} & 2448 \\
2  & \textcolor[rgb]{0,0,1}{1684} & 58                           & \textcolor[rgb]{0,1,0}{1684} & 692                        & \textcolor[rgb]{0.8,0.3,0.3}{1684} & 1684
   & \textcolor[rgb]{0,0,1}{2448} & 3814                         & \textcolor[rgb]{0,1,0}{2448} & 2058                       & \textcolor[rgb]{0.8,0.3,0.3}{2448} & 6325 \\
3  & 3437                         & \textcolor[rgb]{1,0,0}{428}  & \textcolor[rgb]{0,1,0}{1912} & 775                        & \textcolor[rgb]{0.8,0.3,0.3}{1912} & 1912
   & 5489                         & \textcolor[rgb]{1,0,0}{7814} & \textcolor[rgb]{0,1,0}{7935} & 814                        & 5024                               & 4772 \\
4  & \textcolor[rgb]{0,0,1}{1912} & 563                          & 3437                         & 749                        & \textcolor[rgb]{0.8,0.3,0.3}{4039} & 483
   & 5380                         & \textcolor[rgb]{1,0,0}{7773} & 3814                         & 6107                       & 8313                               & 6570 \\
5  & \textcolor[rgb]{0,0,1}{4039} & \textcolor[rgb]{1,0,0}{1684} & \textcolor[rgb]{0,1,0}{4039} & 841                        & 3437                               & 348
   & 187                          & \textcolor[rgb]{1,0,0}{3097} & \textcolor[rgb]{0,1,0}{2049} & 6219                       & \textcolor[rgb]{0.8,0.3,0.3}{7935} & 7773 \\
6  &  58                          & 171                          & 2543                         & 699                        & 2543                               & 414
   & 1546                         & 6301                         & 7413                         & 6010                       & 4523                               & 7814 \\
7  & 1085                         & \textcolor[rgb]{1,0,0}{348}  & 2347                         & 788                        & 2347                               & 4039
   & 1231                         & 1499                         & 2944                         & 777                        & \textcolor[rgb]{0.8,0.3,0.3}{2049} & 7935 \\
8  & 698                          & \textcolor[rgb]{1,0,0}{483}  & 1888                         & 743                        & 2266                               & 428
   & 8301                         & \textcolor[rgb]{1,0,0}{7935} & 6095                         & 6516                       & 7413                               & 3097 \\
9  & 567                          & \textcolor[rgb]{1,0,0}{414}  & 1800                         & 750                        & 1941                               & 376
   & 1805                         & 4173                         & 5489                         & 5020                       & 6017                               & 2049 \\
10 & \textcolor[rgb]{0,0,1}{428}  & \textcolor[rgb]{1,0,0}{376}  & 1663                         & 802                        & 1985                               & 475
   & 2944                         & 5380                         & \textcolor[rgb]{0,1,0}{7773} & 4670                       & \textcolor[rgb]{0.8,0.3,0.3}{7773} & 1763 \\

\bottomrule

\end{tabular}
\label{table_top_10}
\end{table*}

\subsection{SI model}


\replaced{Detailed}{Then, detail} experiments \replaced{were}{should be} carried out to show which measure is more effective and reasonable. In \added{the} SI method, each node has two states: infected \deleted{state} and susceptible \deleted{state}. Some \deleted{initial} nodes \replaced{were initially}{would be} selected as infected nodes. \replaced{A}{The} susceptible node \replaced{becomes}{would be} infected by these \deleted{infected nodes into the infected state}, and \added{it} can no longer return to the susceptible state. \replaced{Additionally}{In addition}, \replaced{the state of each node}{every nodes' state} can only be affected by \replaced{its}{their} neighbor\added{ing} nodes with \added{a} probability $\lambda$. In this \replaced{study}{section}, \added{the} SI model is applied to measure the infection ability of some selected initial nodes\added{,} which is positively correlated with \replaced{the degree of importance of a node}{nodes' importance degree}. The top\added{-}10 nodes obtained by different methods are used as \added{the} initial \replaced{infected}{infectious} node\added{s}, and the rest of nodes are defined as susceptible nodes. \replaced{Every}{In each} time \emph{$t$}, infected nodes have \replaced{a spread}{spreading} rate \emph{$\lambda  = {(1/2)^\beta }$} \replaced{for}{to} infect\added{ing} their neighbor\added{ing} susceptible nodes, and the total number\added{s} of infected \deleted{nodes} and susceptible nodes \replaced{are equal to}{equals to} the number of nodes $\left| N \right|$ in complex networks. $\beta$ has different settings for \added{the} different scales of \added{the} networks. After \added{infection at time} \emph{$t$} \deleted{time infection}, the initial nodes with \added{a} higher \replaced{infection}{infectious} ability \deleted{would} cause \replaced{a greater}{more} number of infected nodes in the network\added{,} which can \replaced{indicate}{show} the importance of these nodes. The number of infected nodes \emph{$F(t)$} \replaced{at}{in} some specific time \emph{$t$} is chosen as an indicator to measure the \replaced{infection}{infectious} ability of \added{the} initial \replaced{infected}{infection} nodes. \replaced{A higher}{More} number of infected nodes \replaced{indicates that the infection}{is, the stronger the infectious} ability of the initial nodes \replaced{is stronger}{are,} and \replaced{that}{more importance of} the initial nodes are \added{more important}.

\begin{figure}[!htbp]
\centering
\mbox{
\subfigure[Jazz network]{\includegraphics[scale=0.5]{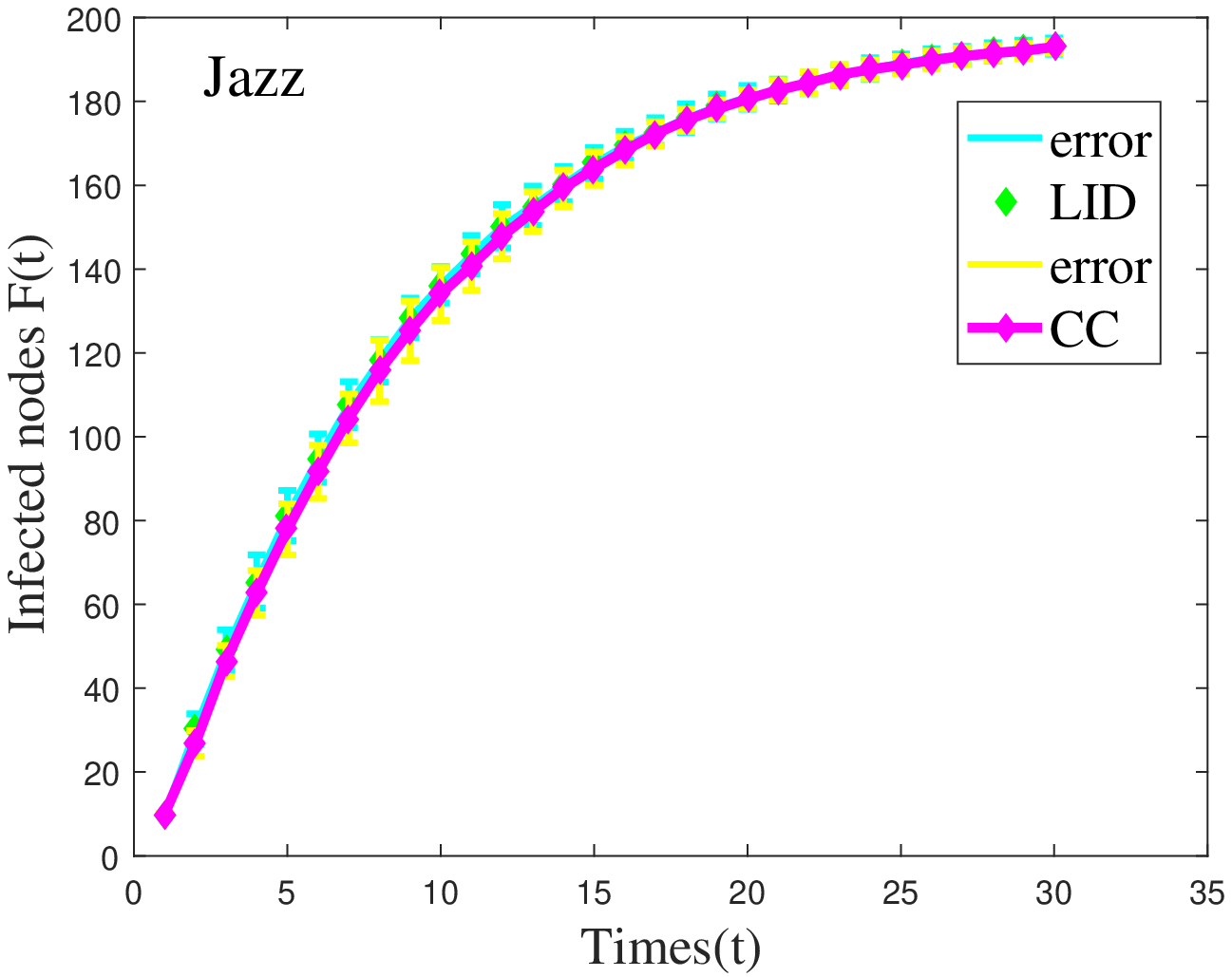}} \quad
\subfigure[Political blogs network]{\includegraphics[scale=0.5]{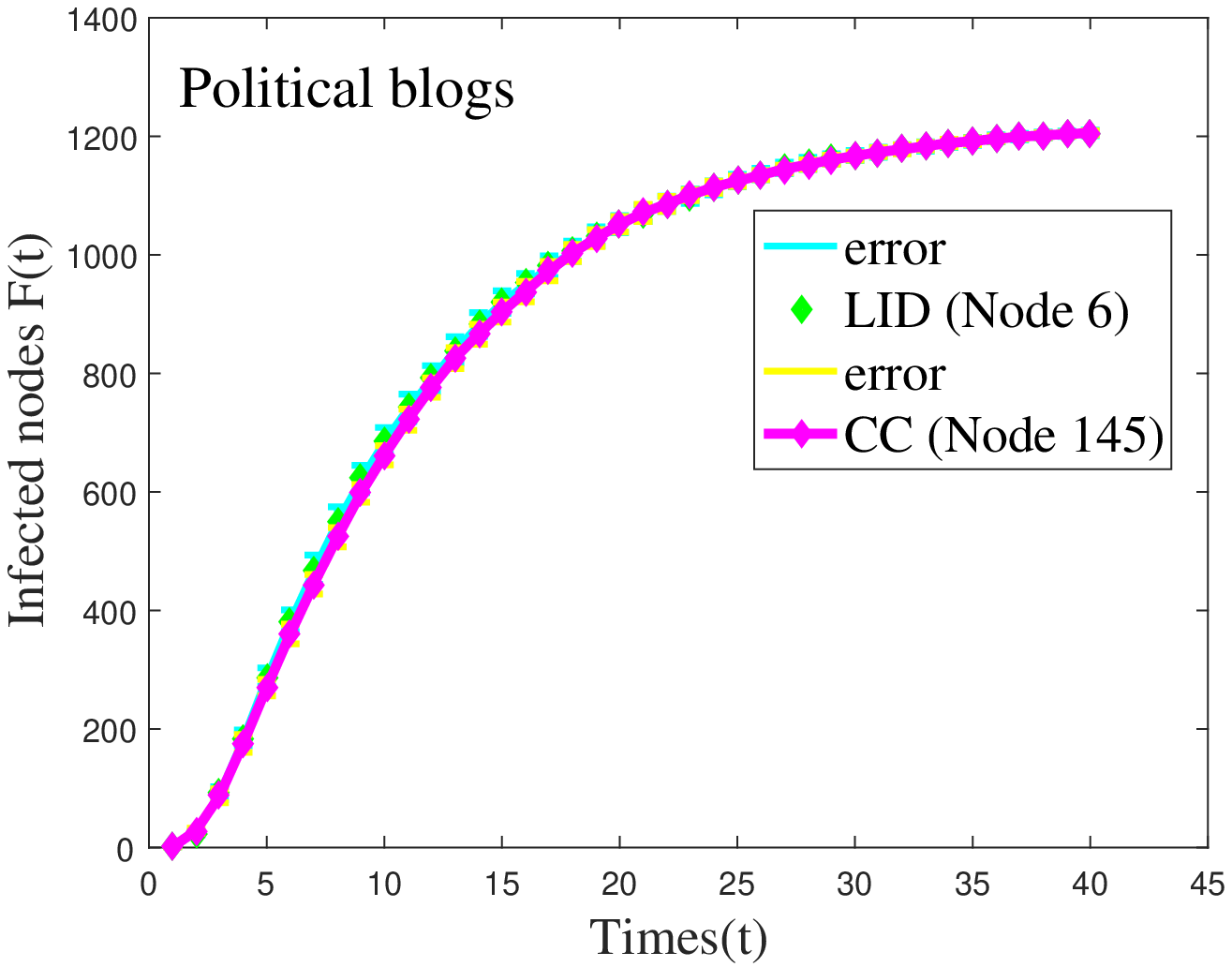}}
}
\mbox{
\subfigure[Facebook network]{\includegraphics[scale=0.5]{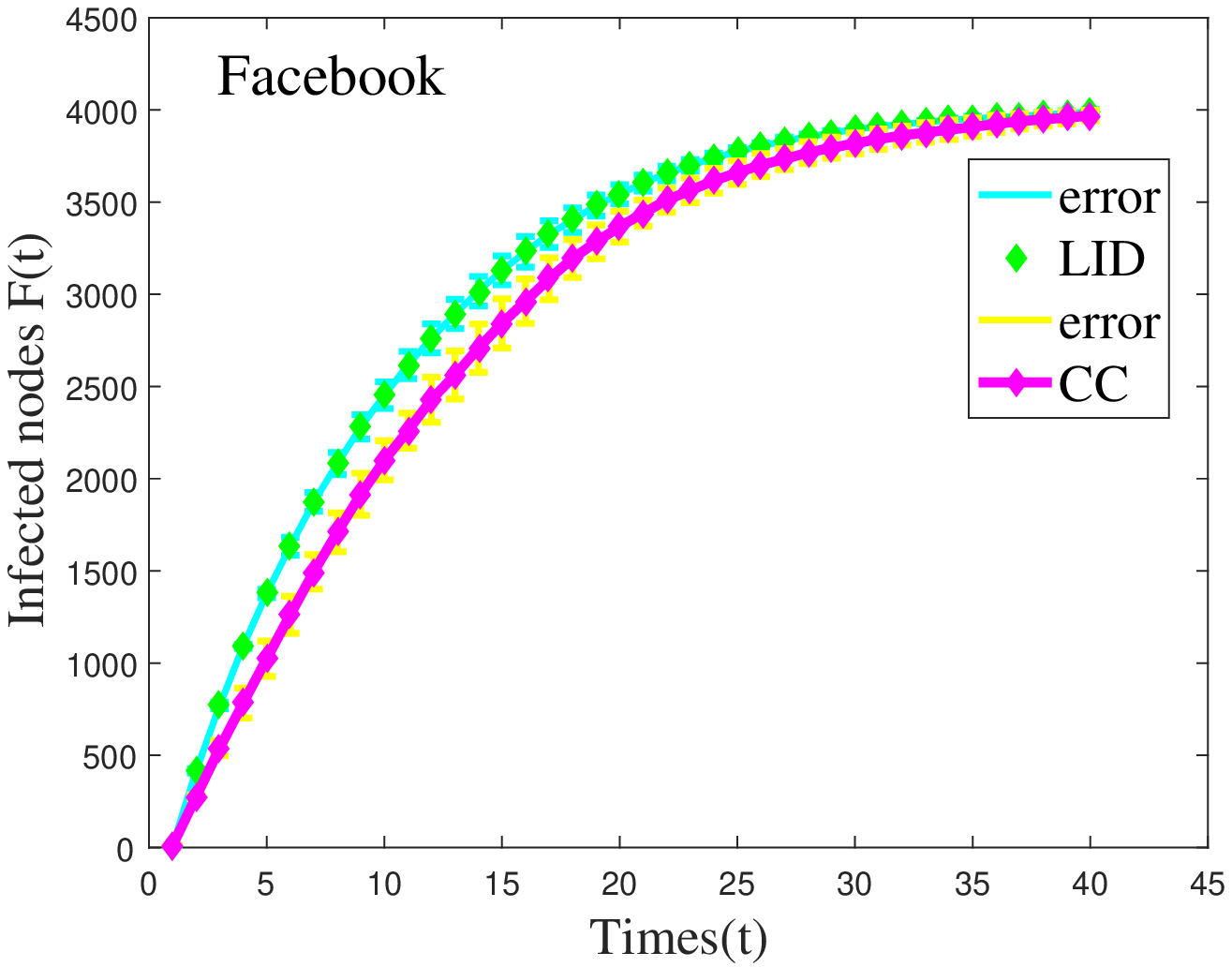}} \quad
\subfigure[Collaboration network]{\includegraphics[scale=0.5]{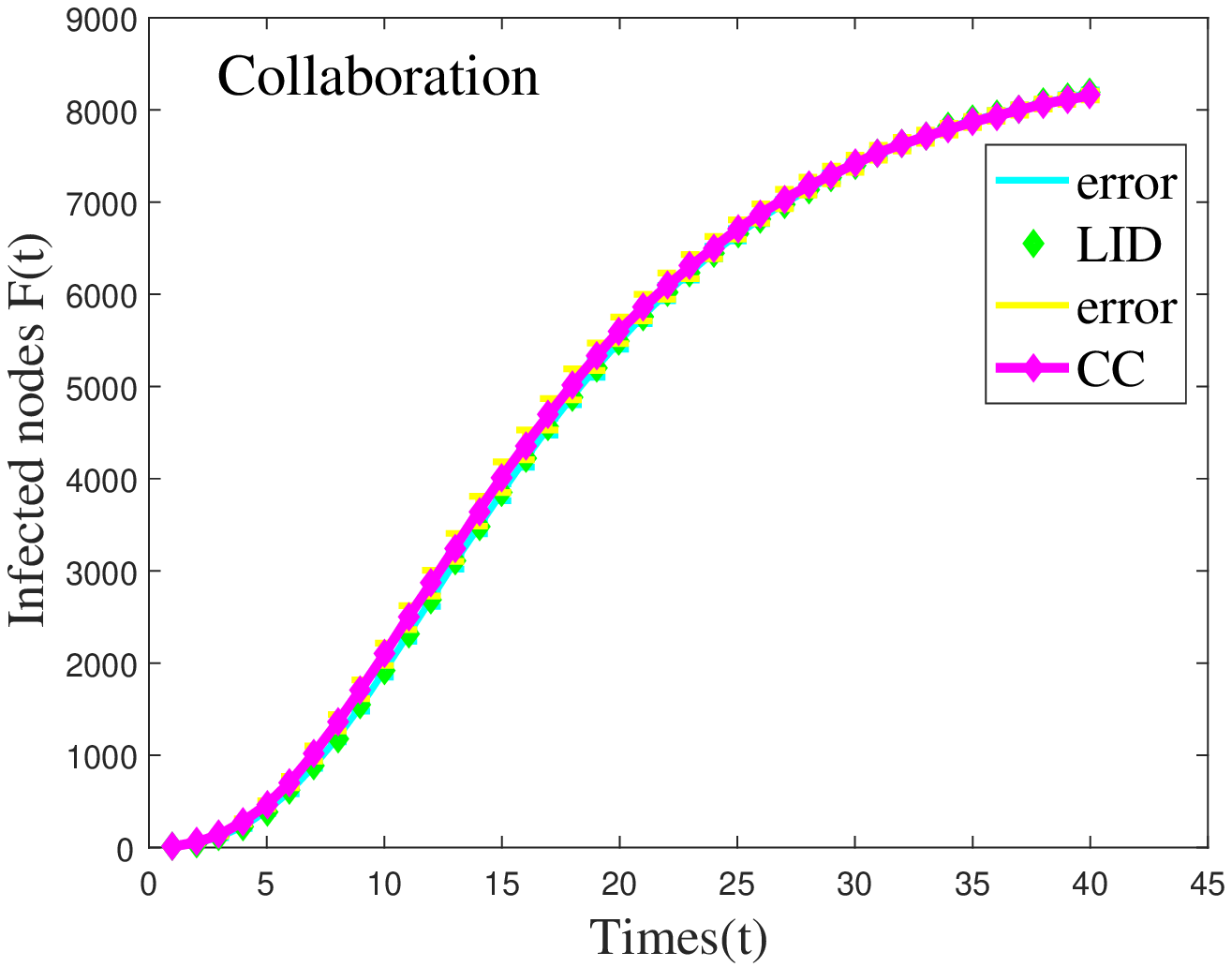}}
}
\caption{\textbf{\replaced{N}{The n}umber of infected nodes \replaced{for}{with} different initial nodes (\replaced{top}{Top}-10 nodes) obtained by \added{the} LID and CC in four networks.} The \replaced{infection}{infectious} ability of \added{the} top-10 nodes \replaced{for the}{in} LID and CC \replaced{is}{are} compared in this figure, and \added{a} higher \added{number of} infected nodes \emph{$F(t)$} \replaced{indicates that the initial nodes have a}{shows} higher \replaced{infection}{infectious} ability\deleted{ of initial nodes}. The results are obtained \replaced{from}{by} 50 independent experiments \replaced{with}{when} $\beta = 3$.}
\label{fig_SI_CC}
\end{figure}

\begin{figure}[!htbp]
\centering
\mbox{
\subfigure[USAir network]{\includegraphics[scale=0.5]{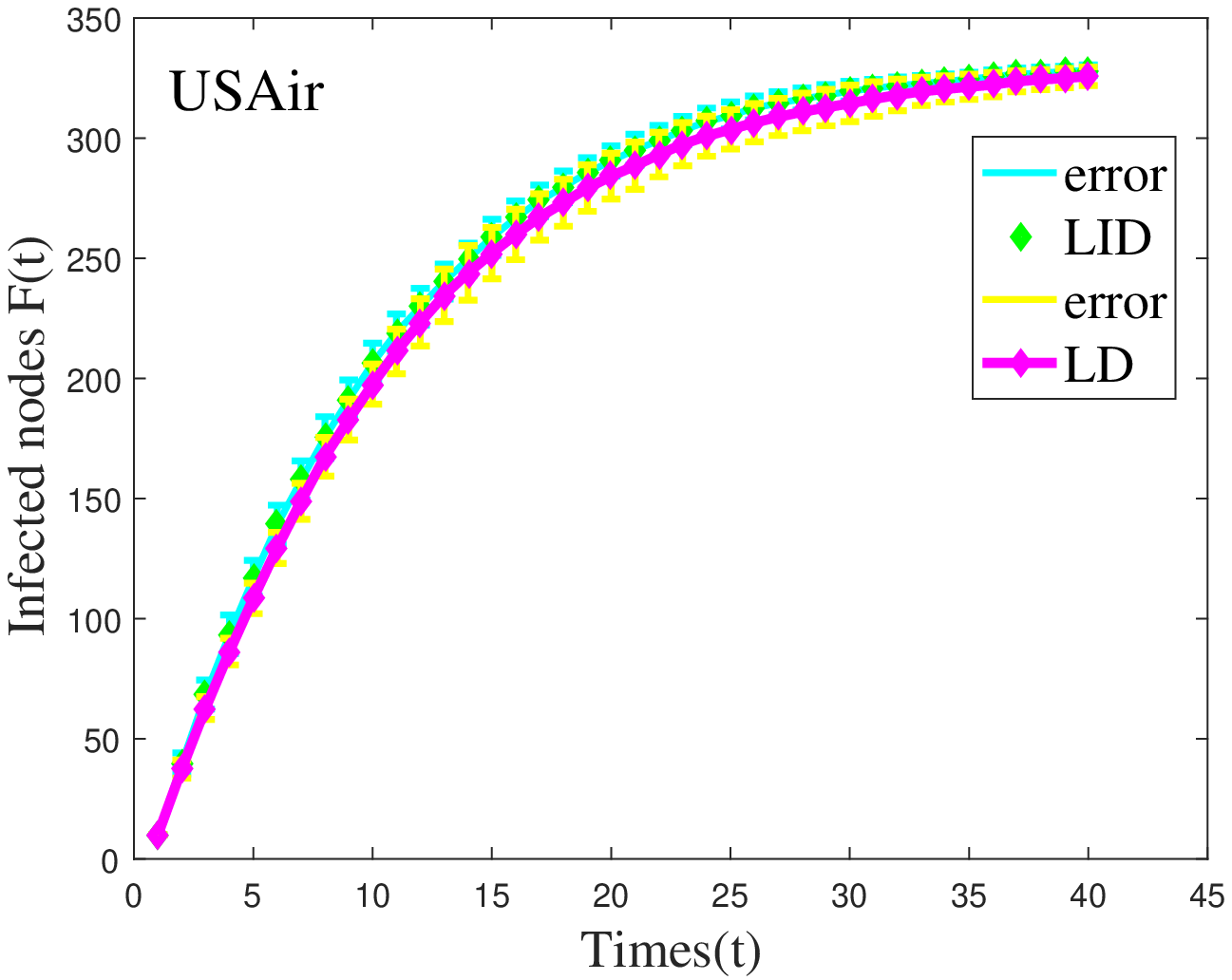}} \quad
\subfigure[Jazz network]{\includegraphics[scale=0.5]{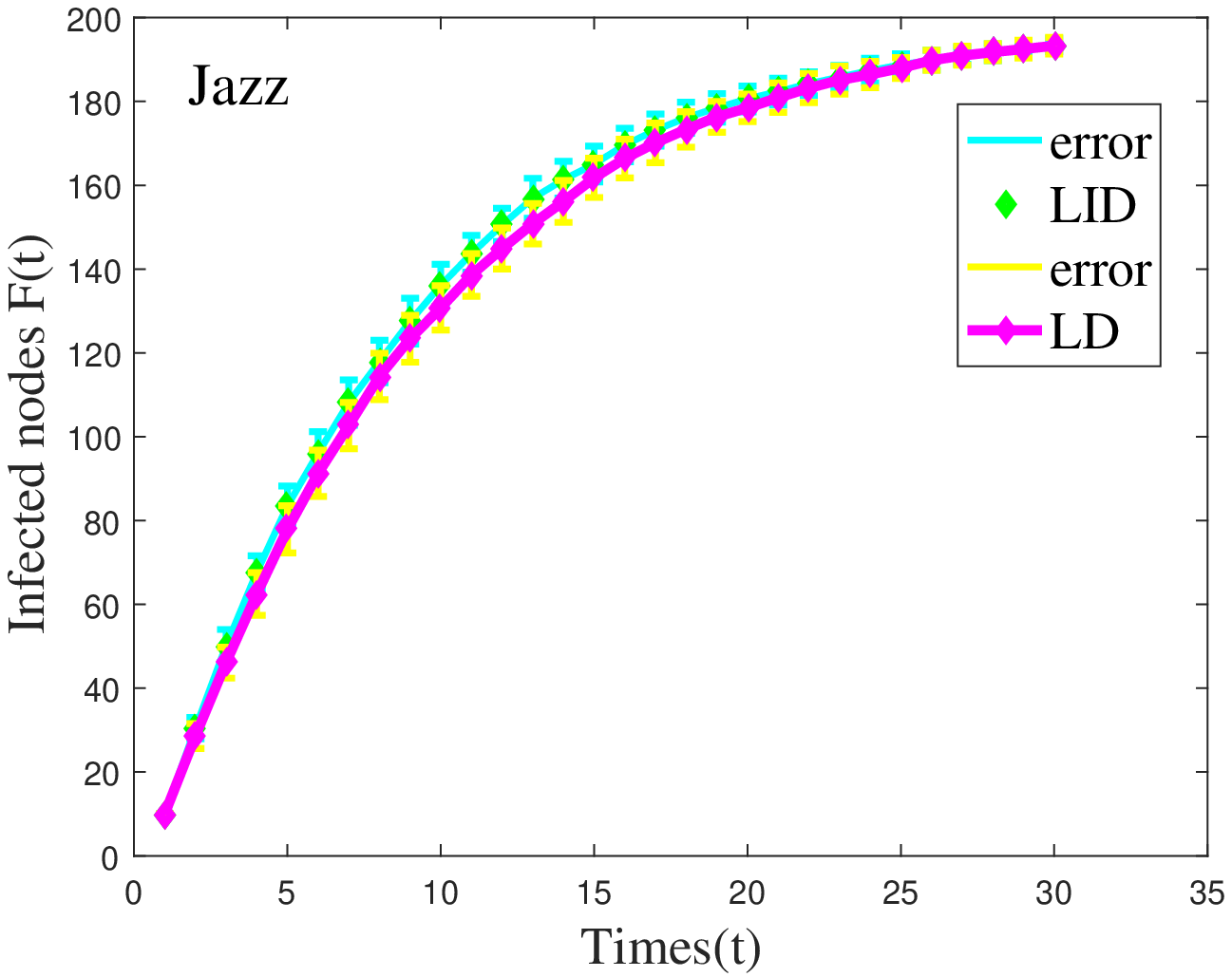}}
}
\mbox{
\subfigure[Karate network]{\includegraphics[scale=0.5]{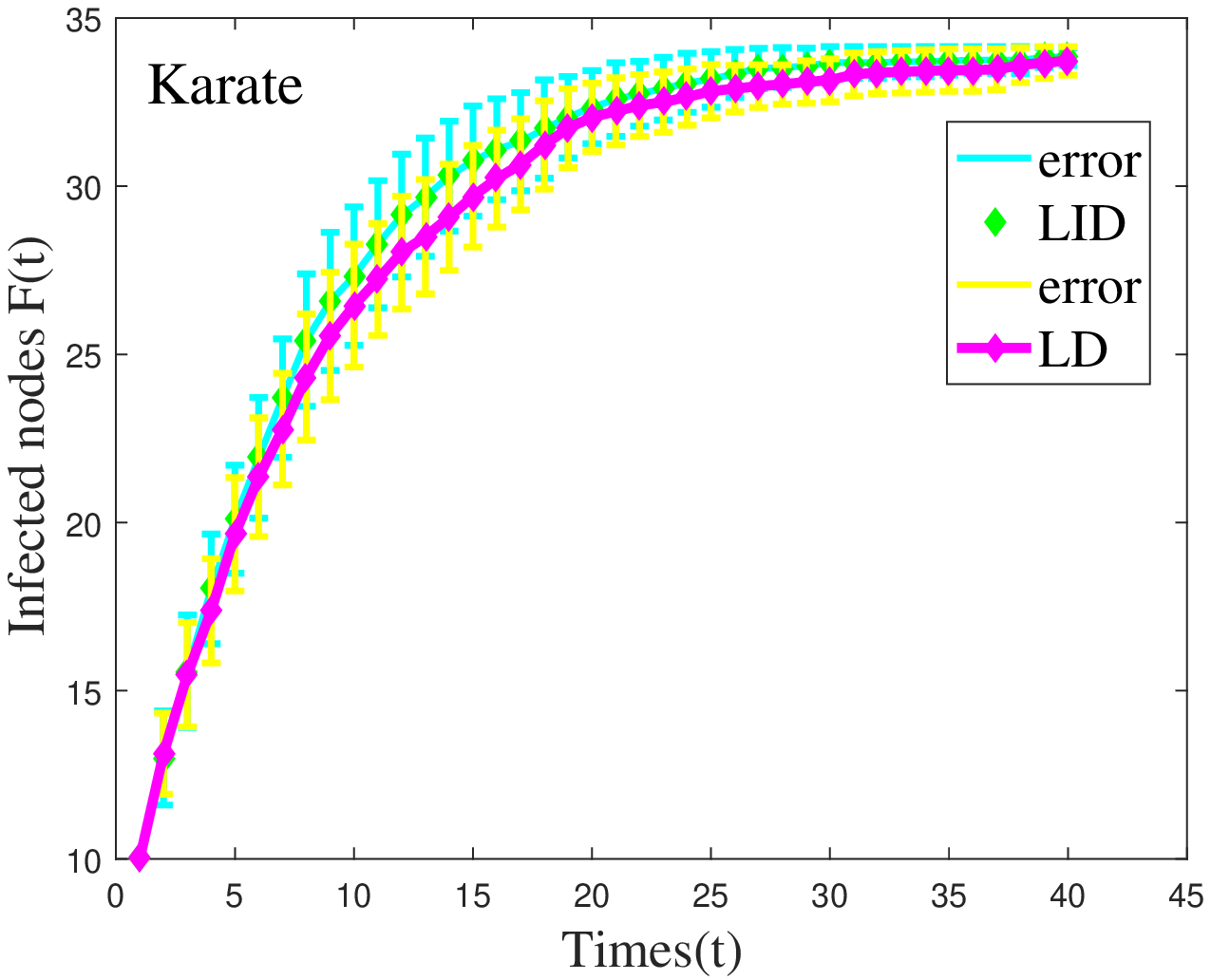}} \quad
\subfigure[Collaboration network]{\includegraphics[scale=0.5]{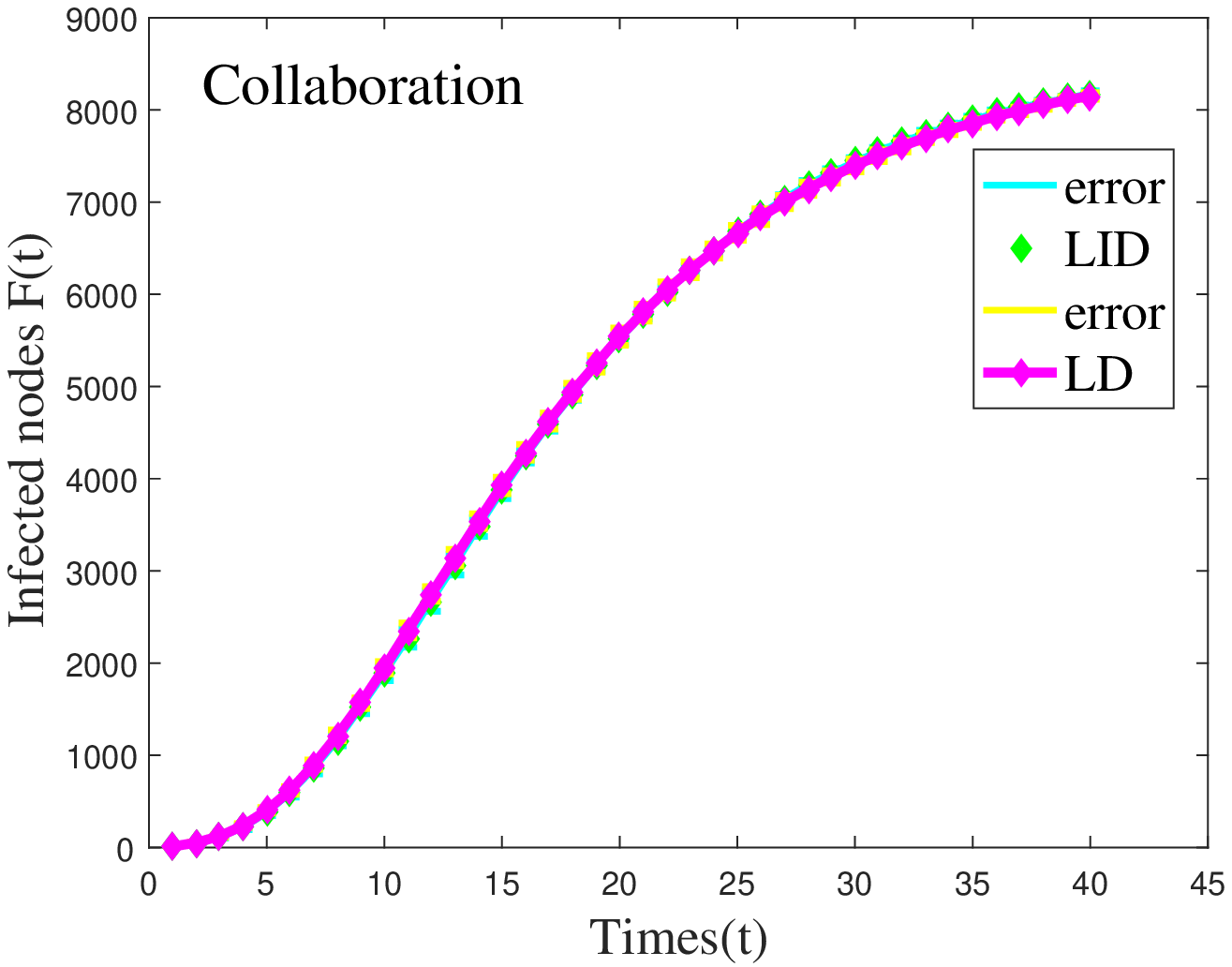}}
}
\caption{\textbf{\replaced{N}{The n}umber of infected nodes \replaced{for}{with} different initial nodes (\replaced{top}{Top}-10 nodes) obtained by \added{the} LID and LD in four networks.} The \replaced{infection}{infectious} ability of \added{the} top-10 nodes \replaced{for the}{in} LID and LD \replaced{is}{are} compared in this figure, and \added{a} higher \added{number of} infected nodes \emph{$F(t)$} \replaced{indicates that the initial nodes have a}{shows} higher \replaced{infection}{infectious} ability\deleted{ of initial nodes}. The results are obtained \replaced{from}{by} 50 independent experiments \replaced{with}{when} $\beta = 3$.}
\label{fig_SI_LD}
\end{figure}

\added{The LD} and CC are selected as comparison methods \deleted{in this section} because of their consideration of information. The \replaced{LID}{proposed method} \replaced{is}{would be} compared with \added{the} CC and LD by \added{the} SI model \replaced{described as follows}{which satisfies the description below}. First\deleted{ly}, the initial \replaced{infected}{infection} nodes are chosen as \added{the} top\added{-}10 nodes obtained by different methods\deleted{,} and \deleted{the detail lists} are \replaced{listed}{shown} in Table \ref{table_top_10}. Then, the infection process \replaced{lasts}{would last} for \added{a time} \emph{$t$}\deleted{time}, and the number of infected nodes \emph{$F(t)$} \replaced{is}{would be} recorded. Lastly, every experiment \replaced{is carried out independently and is}{would be} repeated 50 times \deleted{independently} with \emph{$\beta  = 3$}\replaced{. T}{, and t}he results \replaced{are}{would be} the average of 50 experiments\added{,} which are shown in Fig. \ref{fig_SI_CC} and Fig. \ref{fig_SI_LD}.


\replaced{From}{Observing from} Fig. \ref{fig_SI_CC}, the number of infected nodes \emph{$F(t)$} increases with \added{the} transmission time\deleted{,} and \added{eventually} reaches \added{a} stable value\deleted{finally}. Because the \added{lists of} top\added{-}10 nodes \deleted{lists} obtained by \added{the} CC \replaced{for the}{in} USAir \deleted{network} and Karate network\added{s} are the same \replaced{as that for the}{with} LID, the \replaced{other}{rest of} four networks \replaced{are}{is} used in SI model to compare \added{the} LID and CC. In \added{the} Jazz network, \added{the} LID is slightly better than \added{the} CC, which can be seen \replaced{for \emph{$t$}~=~5--20}{from time \emph{$t$} 5 to 20}. In \added{the} Political blogs network, because \added{the} CC has \replaced{nine of the}{9} same top\added{-}10 nodes \replaced{as the}{with} LID, the \deleted{only} one different node is chosen as \added{the} initial infected node\deleted{s} to simulate \added{the} SI model to show the performance \added{difference} between \added{the} LID and \added{the} CC. The performance of \added{the} LID is better than \added{that of the} CC, \replaced{as}{which can be} observed from the early \replaced{time period}{term} in \added{the} SI model. In \added{the} Facebook network, \added{the} LID is clearly superior to \added{the} CC. \emph{$F(t)$} \replaced{for the}{in} LID is \replaced{larger}{bigger} than \added{that for the} CC \replaced{over}{in} the \replaced{entire}{whole} transmission process, and \added{the} LID reaches \added{a} stable value earlier than \added{the} CC. In \replaced{the collaboration}{Collaboration} network, \added{the} LID has \added{a} slightly lower \emph{$F(t)$} in the early \replaced{time period}{term}, but it keeps up with the growth of \added{the} CC after \replaced{$t=30$}{30 times}. The \deleted{comparison} results \replaced{comparing}{between} the LD and LID \replaced{are}{is} shown in Fig. \ref{fig_SI_LD}. In \added{the} USAir network, \added{the} LID has \added{a} stronger spreading ability than \added{the} LD\deleted{,} because the \deleted{infected} number of \added{infected} nodes obtained by \added{the} LID is \replaced{larger}{bigger} than \added{that of the} LD in the \replaced{middle time period}{medium term}. \replaced{Moreover, the}{And} LID is clearly \deleted{more} superior \replaced{to the}{than} LD \replaced{over the entire}{in whole} progress\added{ion of the} \deleted{in} Jazz network. In \added{the} Karate network, \added{the} LID is obviously better than \added{the} LD for the average number of infected nodes\deleted{,} and \deleted{it is} significantly more stable. In \replaced{the collaboration}{Collaboration} network, \added{the} LD \replaced{has a}{shows} similar effectiveness \replaced{as the}{with} LID\deleted{,} because the\added{ir} curves \deleted{of them are} almost overlap\deleted{ping}. Overall\deleted{ speaking}, \added{the superiority of the LID is obvious in most of the SI experiments from the} \replaced{observations of}{Observing} \emph{$F(t)$} in the comparison between \added{the} CC, LD, and LID \replaced{for}{in} different networks\deleted{, the superiority of LID is obvious in most of SI experiments, }\added{.} \replaced{In}{and in} some cases\added{, the} LID has \deleted{a} similar performance \replaced{as that of}{with} other existing methods\added{,} which only \replaced{have}{has} a slight advantage.

\subsection{\replaced{R}{The r}elationship\added{s} between different measures}

\begin{figure}[!htbp]
\centering
\mbox{
\subfigure[USAir network]{\includegraphics[scale=0.5]{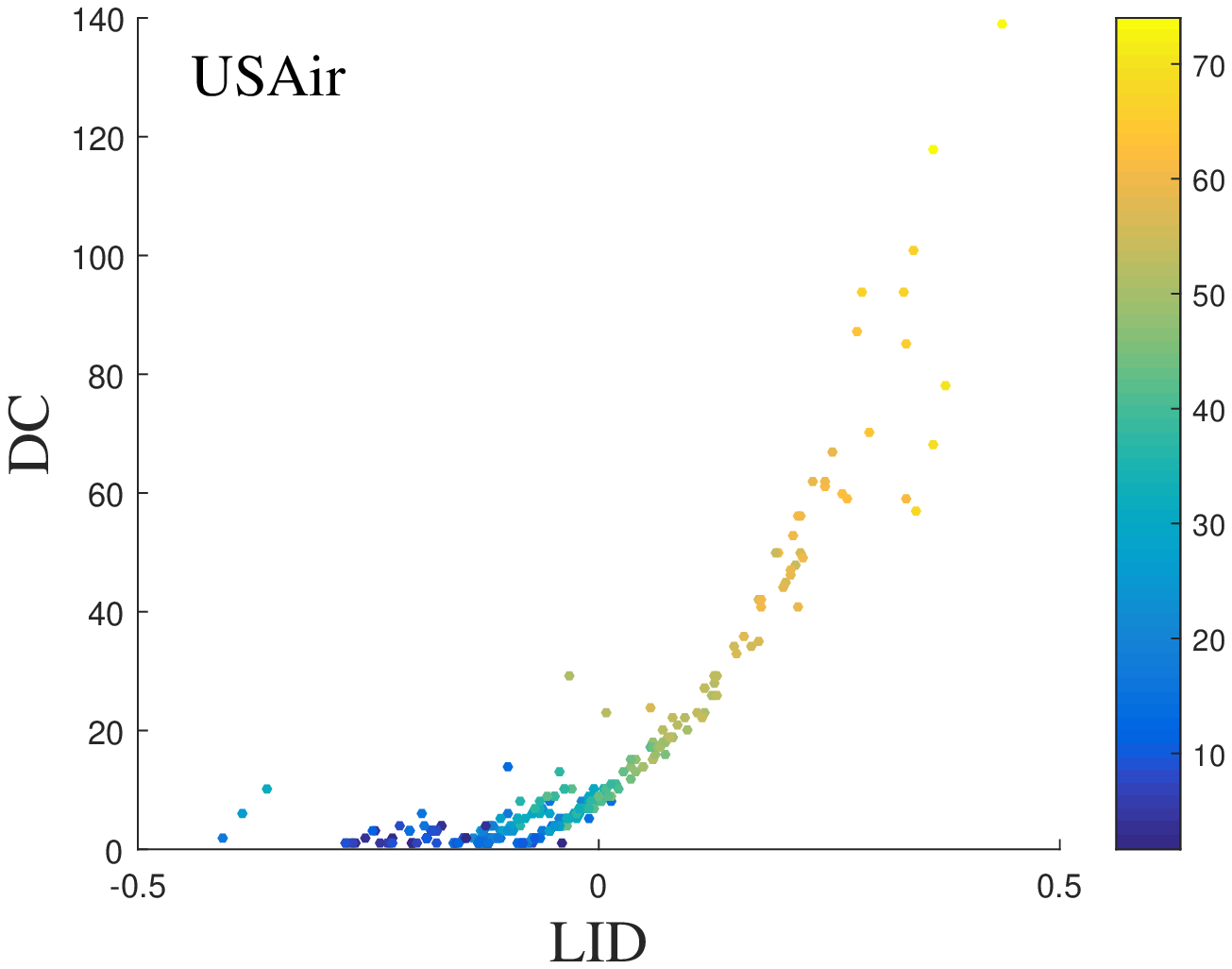}} \quad
\subfigure[Karate network]{\includegraphics[scale=0.5]{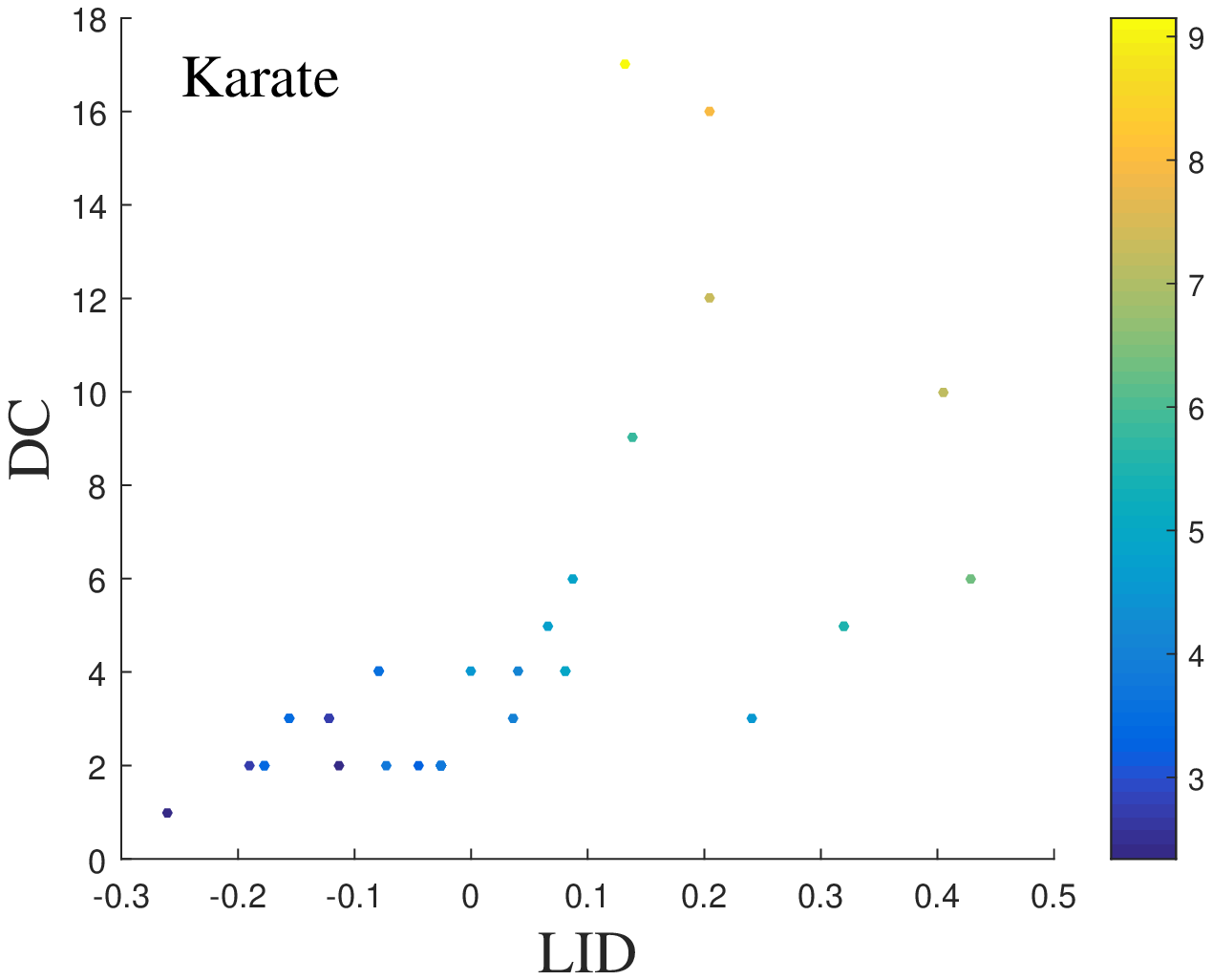}}
}
\mbox{
\subfigure[Political blogs network]{\includegraphics[scale=0.5]{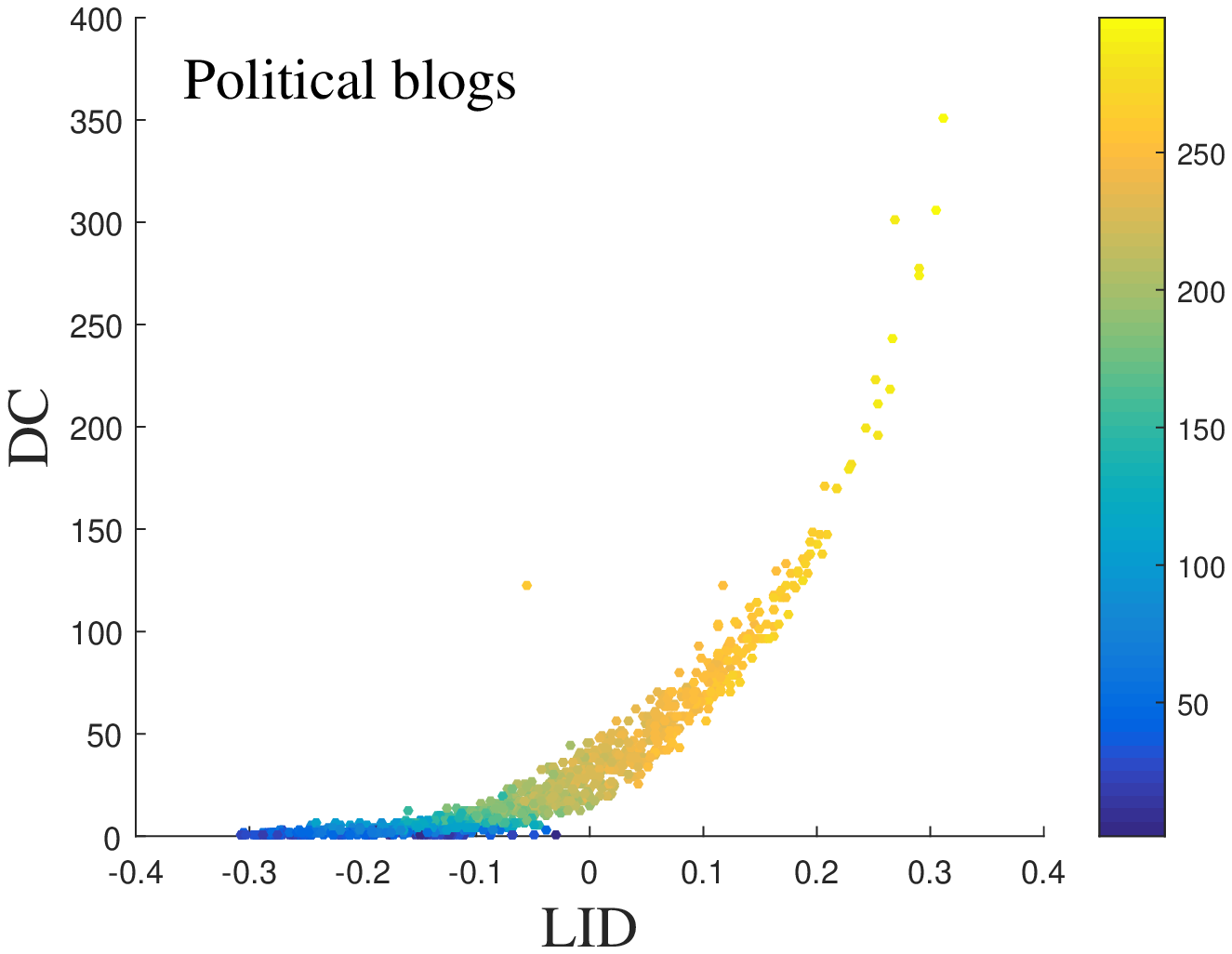}} \quad
\subfigure[Collaboration network]{\includegraphics[scale=0.5]{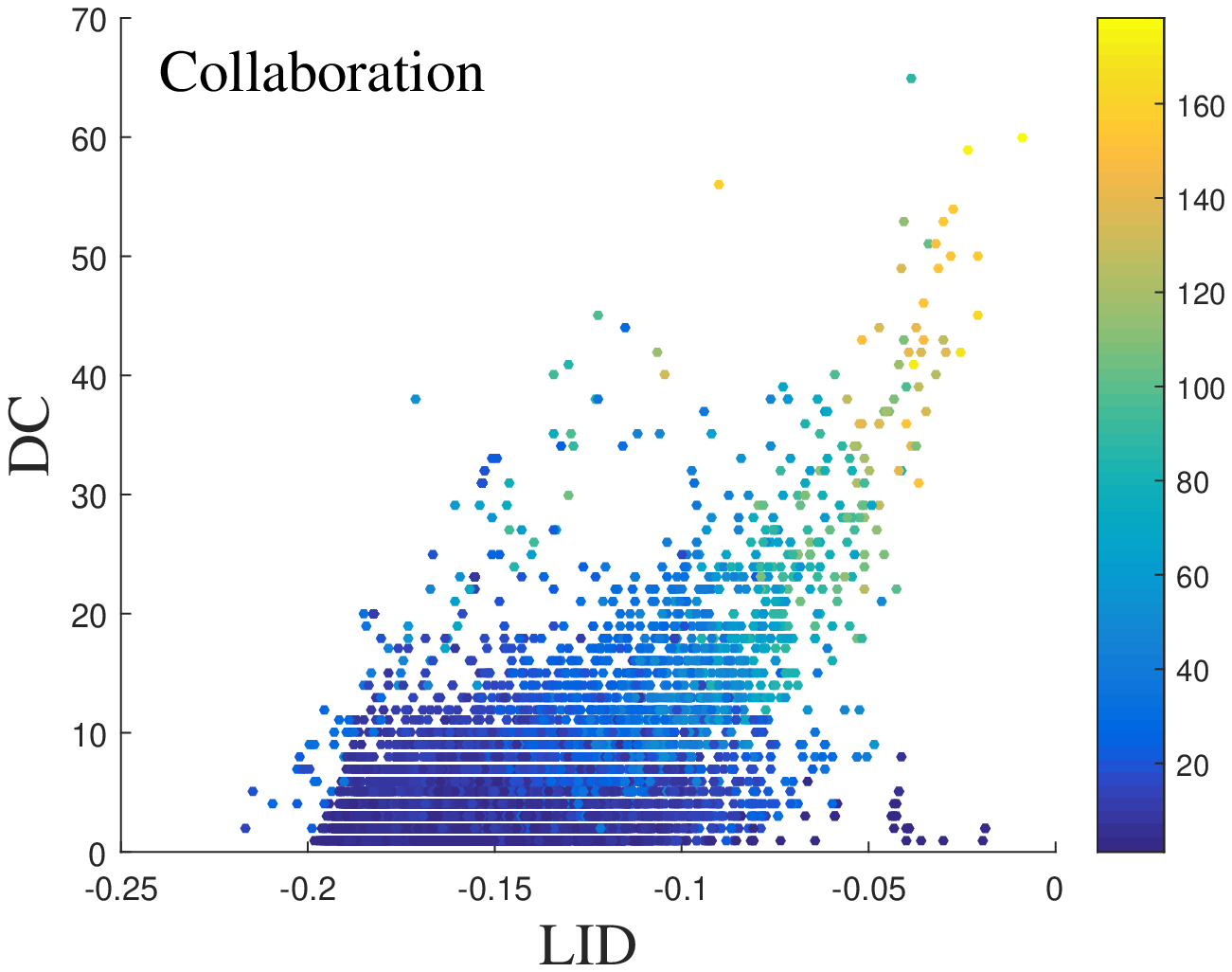}}
}
\caption{\textbf{\replaced{R}{The r}elationship between \added{the} LID and \added{the} DC \replaced{for a}{when} spreading rate $\lambda = 0.05$ in four networks.} Each point represents one node in the network, and the color of \added{a} point represents the number of \deleted{infected} nodes \emph{$F(t)$} \added{infected} with \added{the} selected initial node at $t = 10$, which is obtained \replaced{from}{by} 100 independent experiments. The color and \added{the change in the} value \deleted{change} of \added{the} points show the correlation between \added{the} DC, \added{the} LID, and \added{the} SI model, and the \replaced{monotonic}{monotonous} changes show the similarity between these measures in the general trend.}
\label{fig_corre_DC}
\end{figure}

\begin{figure}[!htbp]
\centering
\mbox{
\subfigure[USAir network]{\includegraphics[scale=0.5]{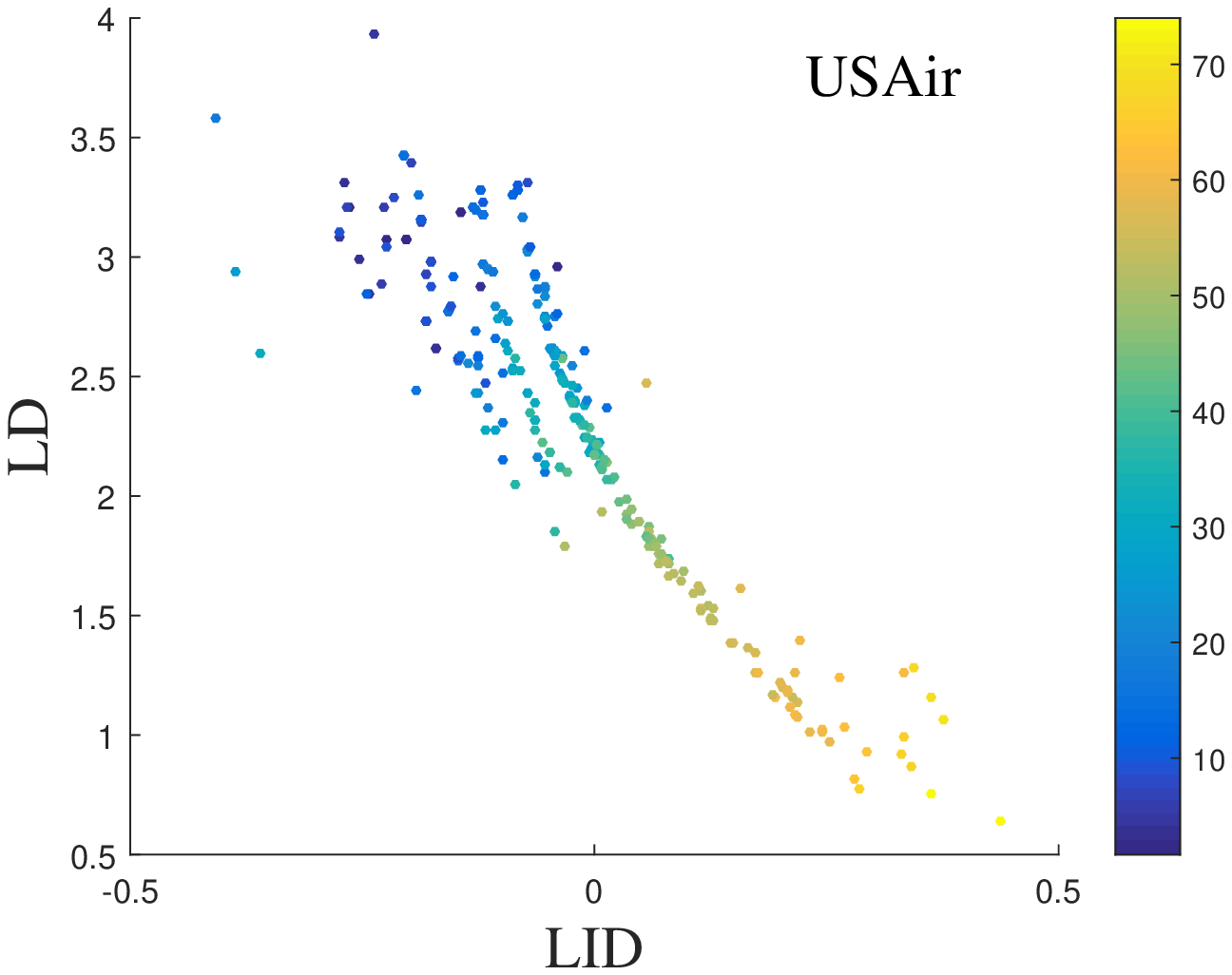}} \quad
\subfigure[Karate network]{\includegraphics[scale=0.5]{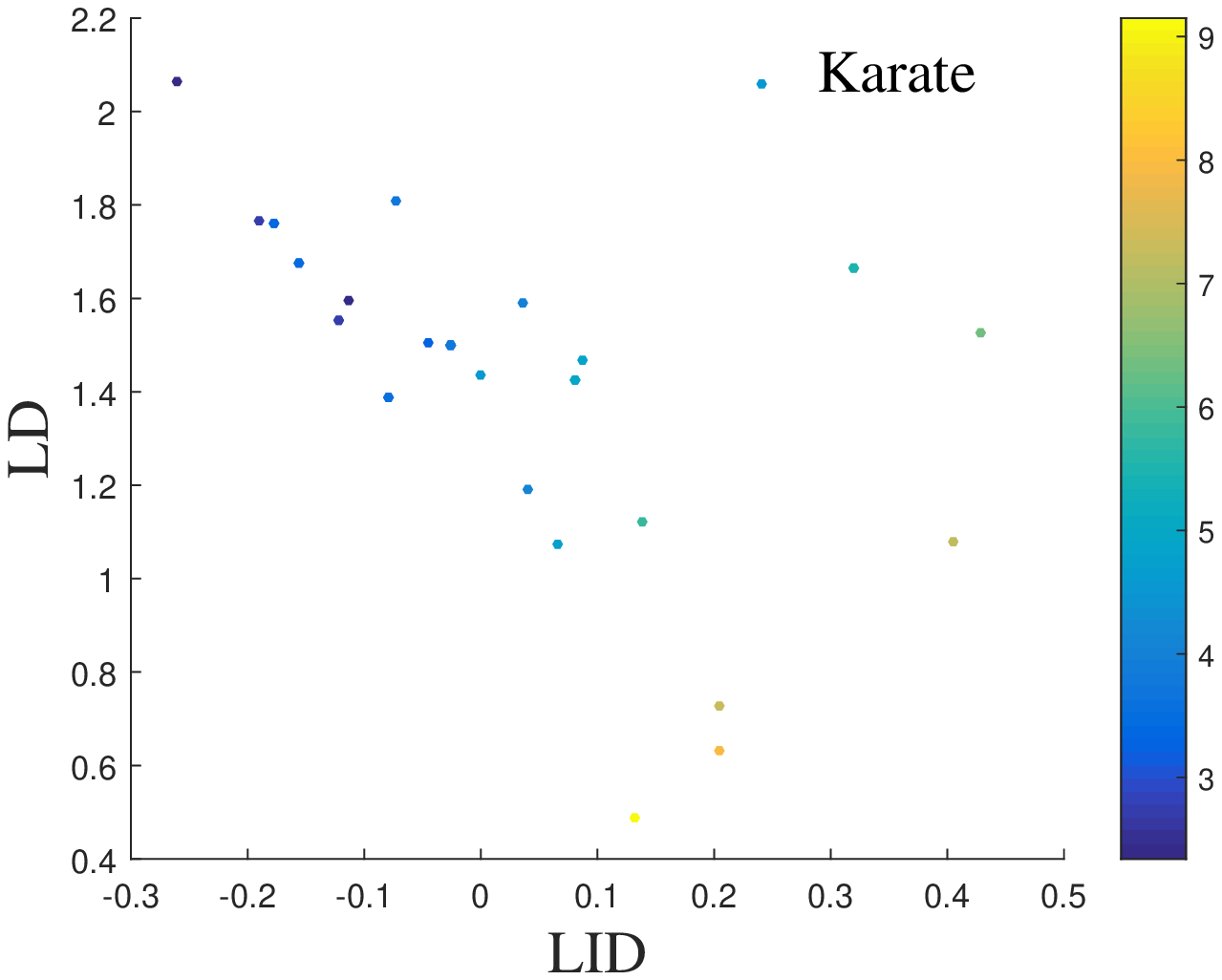}}
}
\mbox{
\subfigure[Political blogs network]{\includegraphics[scale=0.5]{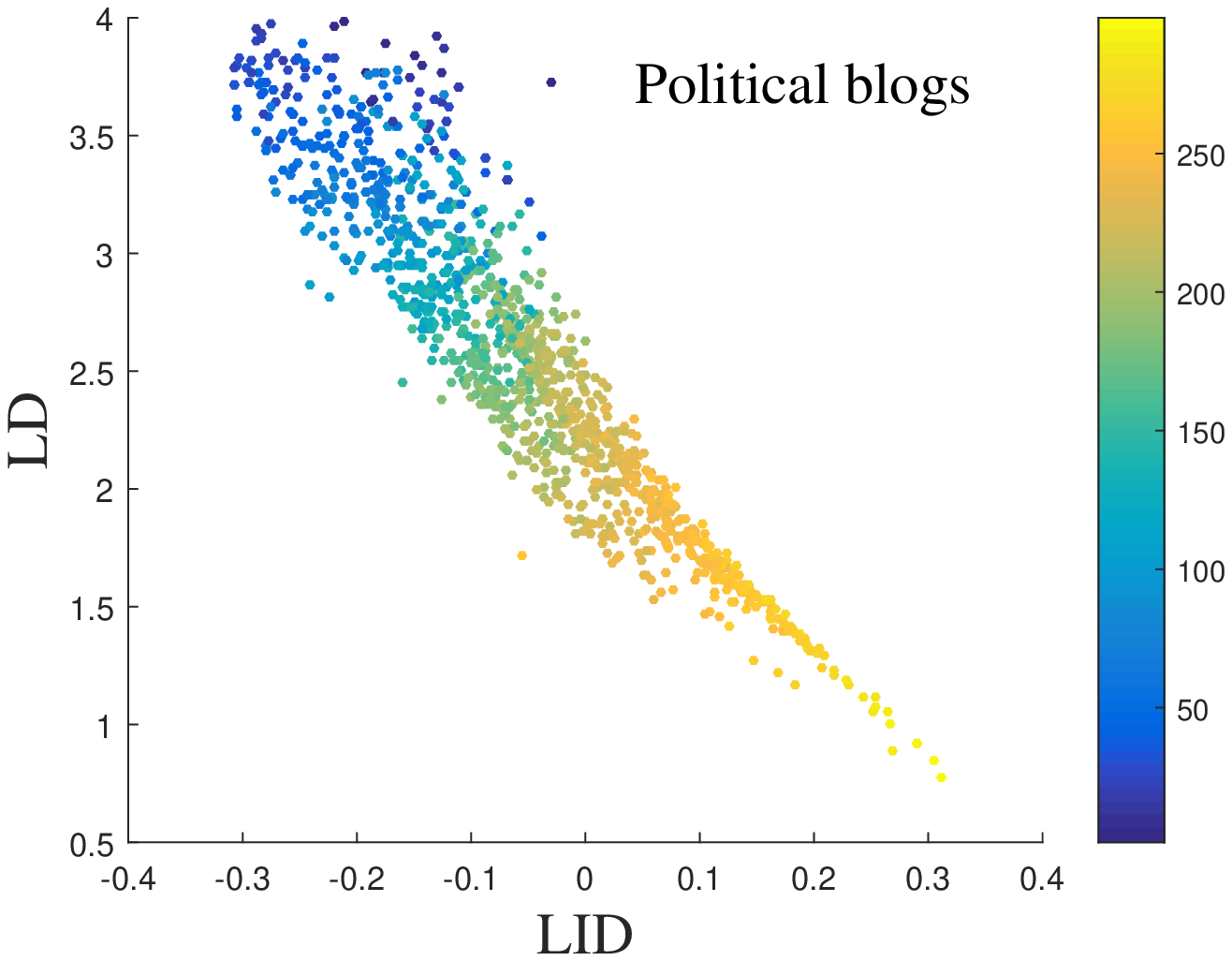}} \quad
\subfigure[Collaboration network]{\includegraphics[scale=0.5]{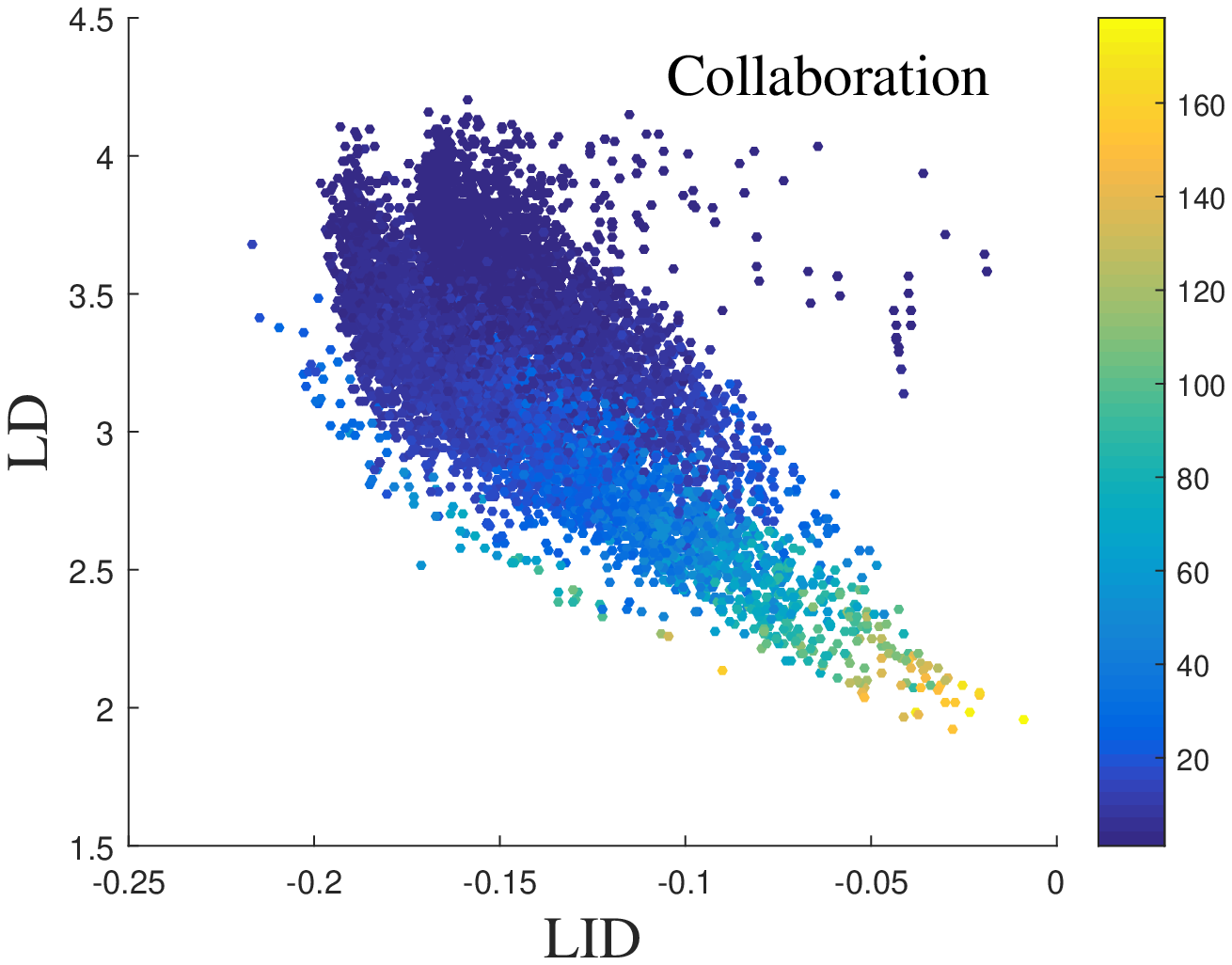}}
}
\caption{\textbf{\replaced{R}{The r}elationship between \added{the} LID and \added{the} LD \replaced{for a}{when} spreading rate $\lambda = 0.05$ in four networks.} Each point represents one node in the network, and the color of \added{a} point represents the number of \deleted{infected} nodes \emph{$F(t)$} \added{infected} with \added{the} selected initial node at $t = 10$, which is obtained \replaced{from}{by} 100 independent experiments. The color and \added{the change in the} value \deleted{change} of \added{the} points show the correlation between \added{the} LD, \added{the} LID, and \added{the} SI model, and the \replaced{monotonic}{monotonous} changes show the similarity between these measures in the general trend.}
\label{fig_corre_LD}
\end{figure}

\replaced{T}{Next, t}o find the relationship between the values obtained by different measures, \replaced{a}{the} relationship graph between different methods is \replaced{presented}{shown in this subsection}. The \deleted{comparison methods are chosen as} DC and LD \added{are chosen for comparison}\deleted{in this subsection}. In the relationship graph, each point represents one node in \added{a} complex network\replaced{.}{s,} \replaced{The}{the} value\added{s} \replaced{along}{of} the horizontal \replaced{and vertical axes}{axis} represent\deleted{s} the LID value of each node\replaced{ and}{,} the \deleted{value of the vertical axis represents the comparison methods (DC or LD)}\added{DC or LD} value of each node, \added{respectively.} \replaced{The}{and the corresponding} color of \added{a} point represents the \replaced{infection}{infectious} ability of \added{the} selected node \replaced{over}{in} 10 \added{time} steps ($F(10)$) when $\lambda = 0.05$ in \added{the} SI model (obtained by 100 independent experiments). When \replaced{a}{the} node \replaced{has a}{with} large LID \replaced{and}{value has} large \replaced{DC or LD}{comparison method's value}, these two methods \deleted{would} have a positive correlation\replaced{;}{,} \replaced{a negative correlation is obtained when a node has a large LID and a small DC or LD}{a negative correlation is the opposite}. The detail\added{ed} results are shown in Fig. \ref{fig_corre_DC} and Fig\added{.} \ref{fig_corre_LD}.

\replaced{From}{Observaing from} the correlation between \added{the} DC and \added{the} LID in Fig. \ref{fig_corre_DC}, the node\added{s} with large LID\added{s} have large DC\added{s}, which means \added{that the} DC is positively correlated with \added{the} LID. \replaced{Owing}{Due} to the \replaced{properties}{property} of \added{the} DC, there \replaced{are many}{would be lots of} nodes with same degree \added{of} centrality\added{,} which can be seen \replaced{in}{from} the figure. Thus, \added{the} LID has an obvious change\added{,} but \added{the change in the} DC\deleted{'s change} is relatively small in the early \replaced{time period}{term}, which demonstrates that \replaced{the important nodes cannot be effectively identified with the small DC}{the nodes' importance with a small DC value cannot be effectively identified.} \replaced{Moreover,}{This is also because} there are \replaced{many}{lots of} nodes with \added{a} small degree \added{of centrality}, which follows the scale-free feature of complex networks. \replaced{Therefore,}{So} this phenomenon shows the superiority of \added{the} LID.

\replaced{From}{In the correlation between LID and LD in} Fig\added{.} \ref{fig_corre_LD}, the relationship between \added{the} LID and \added{the} LD \replaced{has a}{is the} negative correlation\replaced{;}{,} that is, nodes with \added{a} high $F(10)$ have \added{a} high LID but \deleted{have} \added{a} low LD. This is because of the features \cite{Pu2014Identifying} of \replaced{the LD}{local dimension} (\replaced{a node with greater importance has a smaller LD}{the more important the node, the smaller the local dimension}). The relation between \added{the} LID and \added{the} LD is \replaced{similar to}{like} a linear relation\added{,} which \replaced{indicates that}{represents} they would give similar rank lists.

In conclusion, \replaced{the}{this} proposed method is the same as classical measures\replaced{;}{,} \replaced{a larger}{the bigger} value \replaced{for a}{of} measure \replaced{indicates a}{represents the} stronger \replaced{infection}{infectious} ability. \added{The} LID can \replaced{achieve}{keep owning} stable correlative performance with other centrality measures in different real-world complex network\added{s}. In addition, \added{the} LID can \added{more effectively} identify the nodes' importance with \added{a} small degree \added{of centrality}\deleted{more effectively}.

\subsection{\deleted{The}Kendall's tau coefficient}

Kendall's tau coefficient \cite{L2016Vital} has been applied to measure the correlation between \added{the} centrality measures and \added{the} \replaced{infection}{infectious} ability measured by \added{the} SI model \cite{L2016Vital}. Kendall's tau coefficient can measure the correlation between two different variables, and \added{a} higher Kendall'\added{s} tau coefficient \replaced{indicates that}{shows} these two variables are more similar\added{,} which can \replaced{obtain a}{get} more effective result.

The definition of Kendall's tau coefficient \replaced{is as follows}{is shown below}. \replaced{For}{There are} two random variables $A$ and $B$, \deleted{and} their \emph{$i$}th combination is \added{denoted by} $({A_i},{B_i})$. When ${A_i} > {A_j}$ and ${B_i} > {B_j}$ or ${A_i} < {A_j}$ and ${B_i} < {B_j}$ \added{simultaneously} occur\deleted{at the same time}, $({A_i},{B_i})$ and $({A_j},{B_j})$ \replaced{are}{would be} considered concordant. $({A_i},{B_i})$ and $({A_j},{B_j})$ \replaced{are}{would be} considered discordant when ${A_i} > {A_j}$ and ${B_i} < {B_j}$ or ${A_i} < {A_j}$ and ${B_i} > {B_j}$ \added{simultaneously} occur \deleted{at rge same time}. In addition, when ${A_i} = {A_j}$ and ${B_i} = {B_j}$, $({A_i},{B_i})$ and $({A_j},{B_j})$ \replaced{are}{would be} considered neither discordant nor concordant. \replaced{Therefore,}{So the} Kendall's tau coefficient $\tau $ is defined as follows\replaced{:}{,}
\begin{equation}\label{equ_kendall}
\tau  = \frac{{{n_c} - {n_d}}}{{0.5n(n - 1)}}
\end{equation}
where ${{n_c}}$ \deleted{is the number of concordant combinations} and ${{n_d}}$ \replaced{are}{is} the number\added{s} of \added{concordant and} discordant combinations, \added{respectively, and} $n$ is the number of combinations in the sequence. \emph{$\tau = 1$} \replaced{indicates that}{demonstrates} the \deleted{nodes' importance ranking} list \added{ranking the nodes' importance} obtained by different methods is the same as \added{the list ranking} the \replaced{infection}{infectious} ability \deleted{ranking list} obtained by \added{the} SI model\replaced{;}{,} \emph{$\tau = 0$} \replaced{is the opposite case}{means that these lists are completely different}.

\begin{figure}[!htbp]
\centering
\mbox{
\subfigure[USAir]{\includegraphics[scale=0.5]{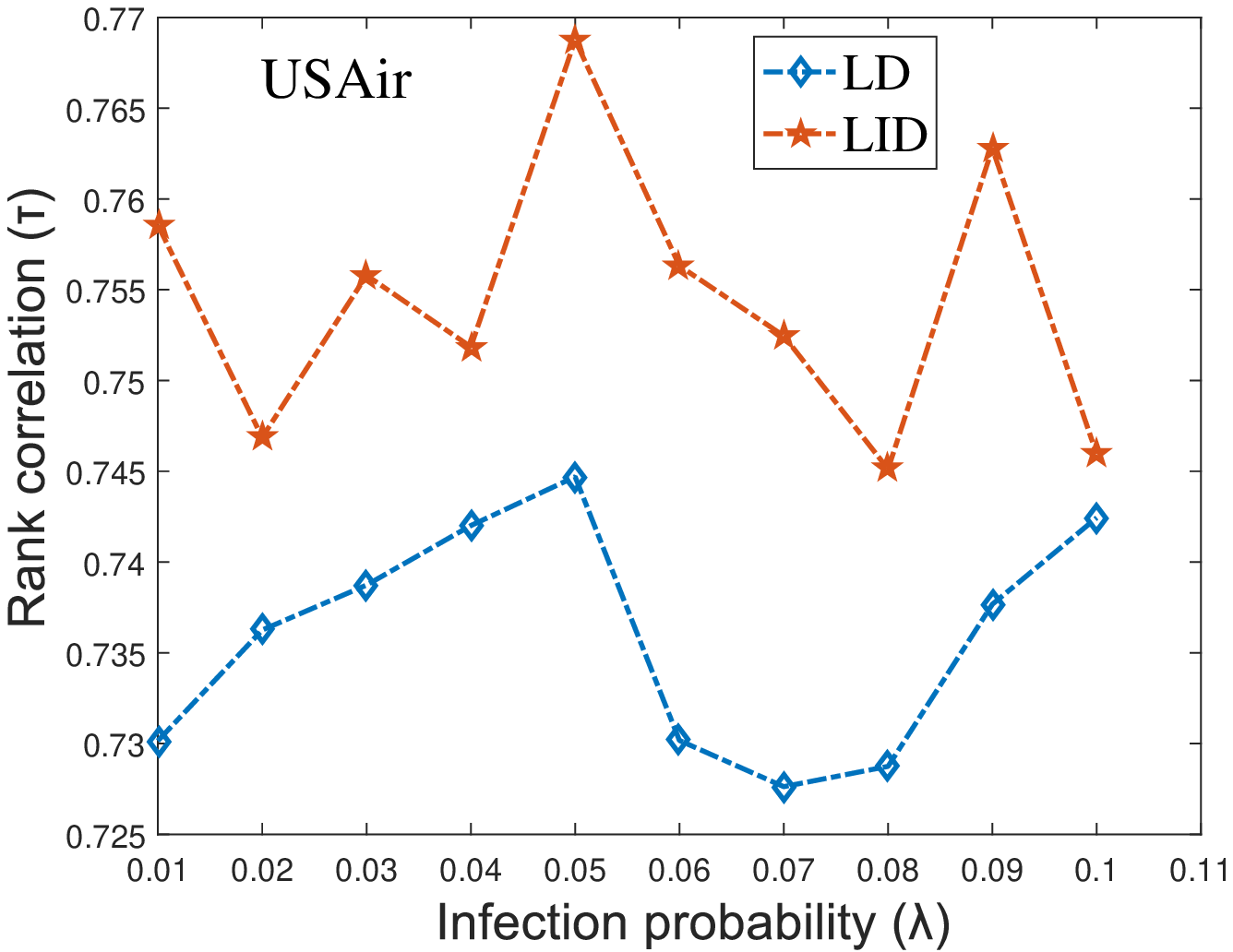}} \quad
\subfigure[Jazz]{\includegraphics[scale=0.5]{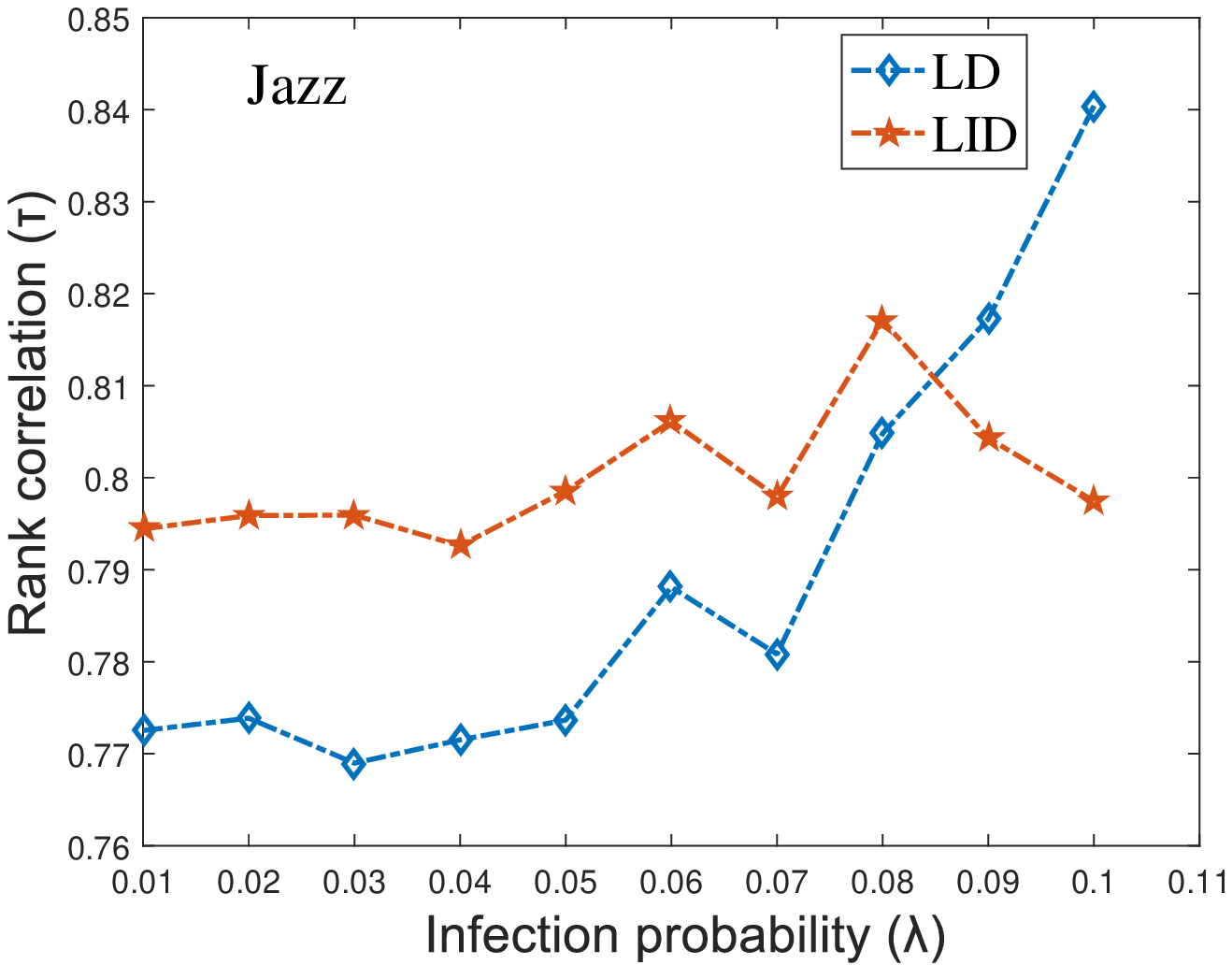}}
}
\mbox{
\subfigure[Karate]{\includegraphics[scale=0.5]{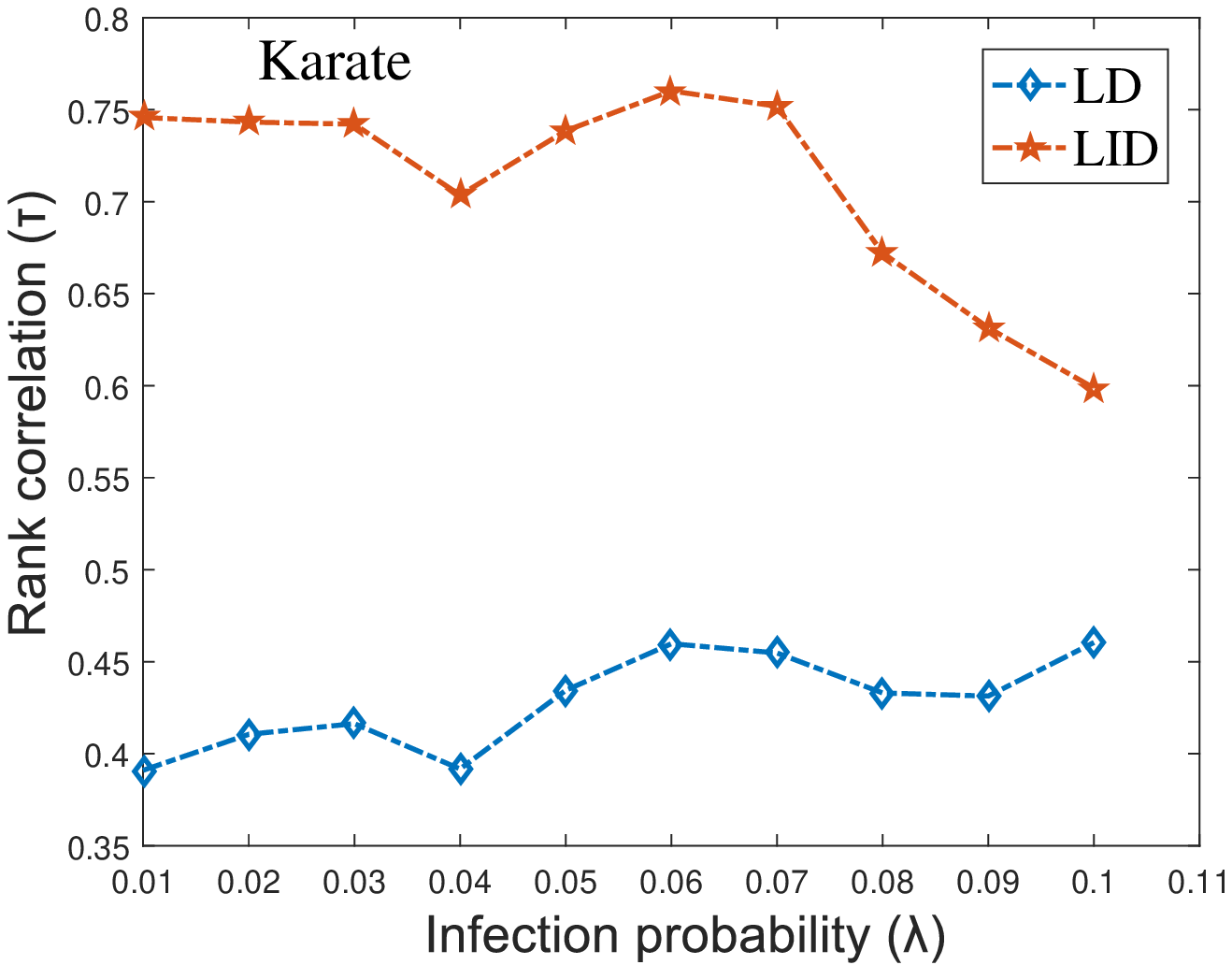}} \quad
\subfigure[Political blogs]{\includegraphics[scale=0.5]{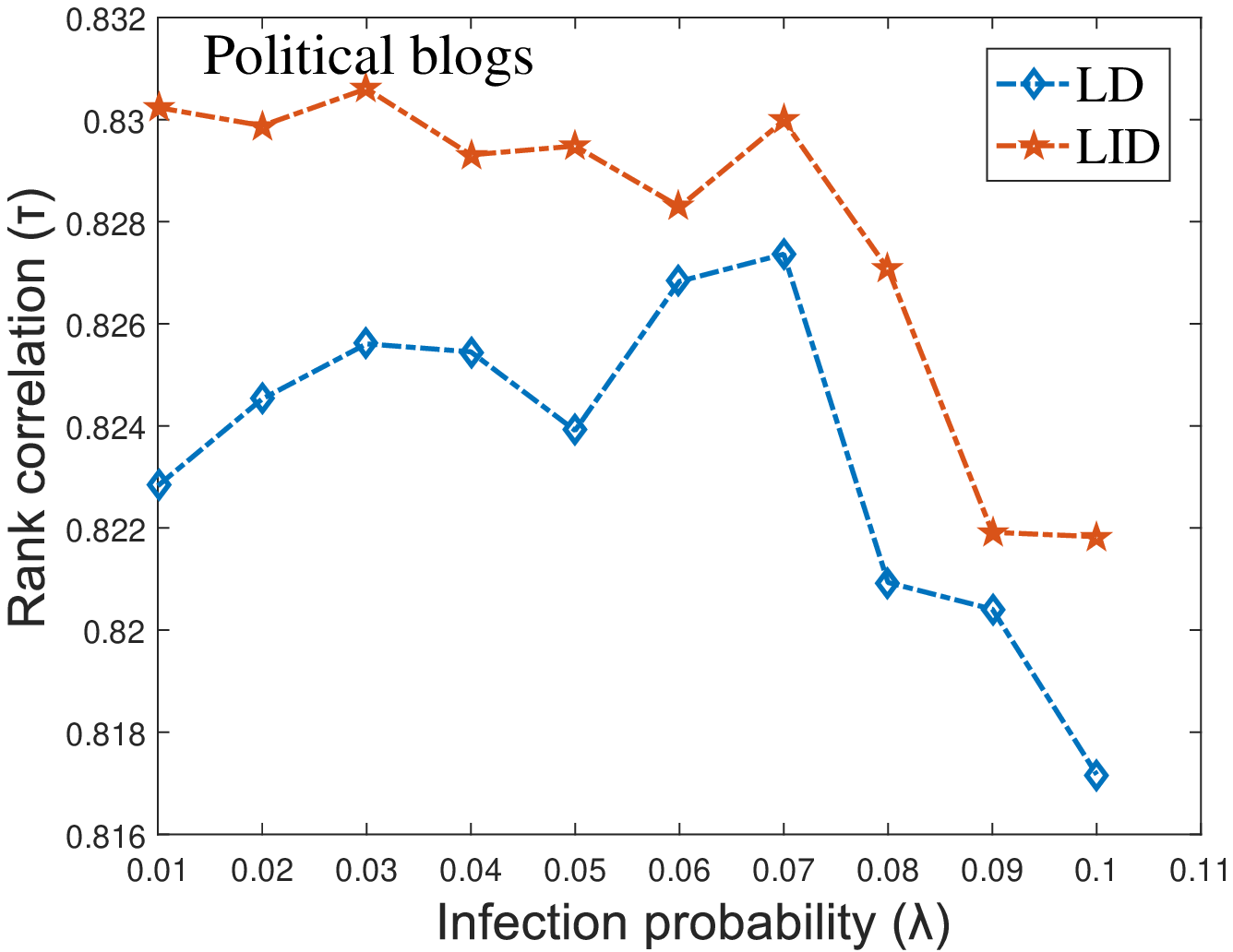}}
}
\caption{\textbf{\deleted{The}Kendall\added{'s tau} coefficient \emph{$\tau $} \added{between the infection ability} obtained by \added{the} SI model and \added{the} LID \replaced{and}{,} LD \replaced{for}{respectively in} four networks.} The \replaced{infection}{infectious} ability of each node is obtained by \added{the} SI model \replaced{from}{in} 100 independent experiments. \replaced{A}{The} higher \deleted{Kendall coefficient} \emph{$\tau $} \replaced{means that a}{represents this } method is more similar \replaced{to the}{with} SI model \replaced{given a}{with the} change \replaced{in the}{of} infection probability \emph{$\lambda$}.}
\label{fig_kendall}
\end{figure}

\deleted{In this section, the }Kendall's tau coefficient $\tau $ between \added{the} LID \added{and LD} and \replaced{the infection}{infectious} ability is compared. The \replaced{infection}{infectious} ability of each node is represented by the number of infected nodes in 10 steps ($F(10)$) \replaced{of the}{in} SI model. \replaced{Additionally}{In addition}, different cases are considered in this experiment. The spreading rate $\lambda$ in \added{the} SI model is \replaced{varied}{considered to} from 0.01 to 0.1 to examine \deleted{the coefficient} $\tau $\deleted{between centrality measure and infectious ability at different spreading rates}. The infection process is \added{independently} repeated 100 times\added{,} \deleted{independently} and \emph{$\tau$} is obtained by averaging. \replaced{A}{The} larger \deleted{the independently} $\tau $ \deleted{is,} \added{indicates that the relationship between the infection ability and the centrality measure is more relevant}

\replaced{The results for}{The} Kendall's tau coefficient $\tau $ \replaced{for}{between LID, LD and infectious ability in} four real-world complex networks are shown in Fig. \ref{fig_kendall}. \replaced{For the}{In} USAir network, \deleted{it can be observed that }\emph{$\tau$} does not have an obvious \replaced{relationship}{rule to change} with the change \replaced{in}{of} $\lambda$, and the difference between \emph{$\tau$} is relative\added{ly} small. In addition, \emph{$\tau$} \replaced{for the}{of} LID is always \replaced{larger}{bigger} than \emph{$\tau$} \replaced{for the}{of} LD, which \replaced{indicates}{represents} the superiority of \replaced{the}{this} proposed method. \replaced{For the}{In} Jazz network, \emph{$\tau$} \replaced{for the}{of} LID is \replaced{larger}{bigger} than \emph{$\tau$} of \added{the} LD when $\lambda$ increases from 0.01 to 0.08\replaced{;}{,} then, \emph{$\tau$} of \added{the} LD \replaced{rapidly increases}{would have a rapid growth} after $\lambda = 0.08$\replaced{, indicating}{which represents} that \emph{$\tau$} of \added{the} LD
\replaced{is larger}{would be bigger} than \emph{$\tau$} of \added{the} LID when $\lambda = 0.09, 0.1$. \replaced{Hence, the}{So this proposed method} performance\deleted{s} \added{of the proposed method is} better \deleted{in} most of the time. \added{The values of} \emph{$\tau$} \replaced{for the}{in} Karate \deleted{network} and Political blogs network\added{s} \replaced{exhibit a}{are a} downward trend. \replaced{For the}{In} Karate network, \emph{$\tau$} of \added{the} LID is much \replaced{larger}{bigger} than \emph{$\tau$} of \added{the} LD, \replaced{indicating that the}{which represents this} proposed method outperforms \added{the} LD. \replaced{For the}{In} Political blogs network, the \replaced{difference in the values of}{different between} $\tau$ \added{for the LID and LD} is small\added{,} but \emph{$\tau$} of \added{the} LID is always \replaced{larger}{bigger} than \emph{$\tau$} of \added{the} LD, which shows the superiority of \added{the} LID. In conclusion, \deleted{the coeffieient }$\tau $ between \added{the} LID and \added{the} \replaced{infection}{infectious} ability is \replaced{larger}{lager} than \deleted{the coefficient} $\tau $ between \added{the} LD and \replaced{the infection}{infectious} ability in most cases. This means \added{that} the result\added{s} obtained by \replaced{the}{this} proposed method \replaced{are}{is} more \replaced{relevant}{related} \replaced{to the}{with} classical \replaced{infection}{infectious} ability, and \added{the} LID can \replaced{maintain}{keep} relatively stable correlative performance than \added{the} LD in most \deleted{of} real-world complex networks. Thus, \replaced{the}{this} proposed method is more effective \replaced{for}{to} identify\added{ing} influencers from this perspective.

\subsection{\replaced{T}{The t}ime consumption}

\replaced{The}{Lastly, the} time consumption of different measures \replaced{for}{in} different networks is \replaced{presented}{recorded in this subsection}. \replaced{All}{We conduct all of these} centrality measures \replaced{were calculated using}{by} MATLAB 2016a on a \added{personal computer} (PC) \added{equipped} with an \replaced{Intel Core}{Inter (R) Core (TM)} i7-5500U \added{central processing unit} (CPU) \added{operating at} 2.40 GHz \deleted{CPU} and 8 GB \added{of random access memory} (RAM). The method with \added{a} lower running time has \added{a} lower computational complexity. The running times of these measures are \replaced{listed}{shown} in Table \ref{table_time}. \replaced{Based on}{Observing from} the result\added{s}, \added{the} DC has the lowest computational complexity\deleted{, whether} on large-\deleted{scale} or small-scale networks. In contrast, \added{the} BC \replaced{has}{runs} the \replaced{highest running}{longest} time and far exceeds \added{those of the} other methods. \replaced{The}{In the rest of the other methods} running time of \added{the} LID is \replaced{approximately}{mostly} half of \added{that of the} LD\deleted{,} and \deleted{in some cases the running time is} less than half \added{in some cases}. This is because \added{the} LID considers the information of nodes (\replaced{quasilocal}{quasi-local} information) whose distance from the central node is less than half of the maximum value of \added{the} shortest distance, but \added{the} LD considers all of the nodes in the network. The running time of \added{the} EC is small \replaced{for}{in} small-scale network\added{s,} but it \replaced{rapidly increases for}{grows fast in} large-scale network\added{s} (\added{as} seen from the comparison with \added{the} LD). \replaced{Additionally}{In addition}, the running time of \added{the} EC is 2\replaced{--}{ to }10 times \added{that of the} LID, which means \added{that the} LID has \added{a} relatively low computational complexity. The reason why the running time of \added{the} CC is smaller than \added{that of the} LID is that \added{the} CC only needs to sum the shortest distance from the central node to other nodes. In conclusion, \added{the} LID \replaced{has a}{runs at a} lower \added{running} time than most other methods\added{,} \replaced{implying that}{which means} the LID reduces the computational complexity.

\begin{table}[!htbp]
\centering
\caption{\textbf{Running time\added{s (in seconds)} of different centrality measures \replaced{for}{in} different real-world network\added{s}.\deleted{(Seconds)}}}
\resizebox{\textwidth}{!}{
\begin{tabular}{ccccccc}
\hline
Network & Time(BC) & Time(CC) & Time(DC) & Time(EC) & Time(LD) & Time(LID)\\
\hline
USAir              & 31.7043     & 0.0033 & 0.0007 & 0.0171   & 0.0201  & 0.0122  \\
Jazz               & 11.1378     & 0.0016 & 0.0006 & 0.0086   & 0.0199  & 0.0058  \\
Karate             & 0.1748      & 0.0004 & 0.0003 & 0.0009   & 0.0017  & 0.0006  \\
Political blogs    & 611.2053    & 0.0435 & 0.0056 & 0.4676   & 0.2551  & 0.1252  \\
Facebook           & 18635.5697  & 0.4985 & 0.0478 & 14.3498  & 2.9104  & 1.2109  \\
Collaboration      & 291827.8567 & 4.3902 & 0.2290 & 193.3321 & 23.7418 & 11.5581 \\
\hline
\end{tabular}}
\label{table_time}
\end{table}

\section{Conclusion}

In this paper, the influencers in complex networks are identified by the \replaced{LID}{local information dimension}. The size of the box covering the central node grows from \replaced{one}{1} to \emph{$ceil({d_i}/2)$}, and the number of nodes within the box is considered \replaced{using the}{by} Shannon entropy\added{,} which can measure the information in the box. Then, the \replaced{LID}{local information dimension} of the central node can be obtained by the correlation between the box information and the size of box. Finally, the \deleted{influential} ability of nodes \added{to influence others} can be ordered \replaced{according to}{by} the value of \added{the} LID. Because of the \added{rule governing the increase in the size of the} box\deleted{size growing rule}, \replaced{the}{this} proposed method considers the \replaced{quasilocal}{quasi-local} information around the central node and reduces the computational complexity. \replaced{Compared}{Comparing} with \replaced{existing}{some exiting} measures \replaced{for}{in} real-world networks, \replaced{the}{this} proposed method is more effective and reasonable, and \replaced{experimental}{the} results \deleted{of the experiments show the} \added{demonstrate its} superiority\deleted{of this method}.

However, \replaced{it can}{this proposed method is still potential to} be improved. One inevitable problem is how to identify \deleted{two spreaders' influential} \added{the} ability \added{of two spreaders to influence others} when they have equal \replaced{LIDs}{value of local information dimension} \emph{${D_{I\_i}}$} (or other measures' results). In future research, the information in the box can be considered \replaced{to be}{into} \replaced{multiscale,}{multi-scale} which can achieve adequate consideration of information, and a better result can be obtained. Therefore, the framework of \added{the} dimension-based approach would be significantly improved \replaced{for}{to} identify\added{ing} \deleted{the} influencers in complex networks.


\section*{Acknowledgment}
The authors greatly appreciate the \replaced{reviewers'}{reviews'} suggestions and \deleted{the}editor's encouragement. \replaced{This}{The} work is partially supported by \added{the} National Natural Science Foundation of China (Grant Nos. 61973332, 61573290, \added{and} 61503237).


\bibliographystyle{plain}
\bibliography{myreference}

\end{document}